\documentclass[aps,prb,twocolumn,groupedaddress,amsmath,amssymb]{revtex4}

\usepackage{graphicx}
\usepackage{subfigure}
\usepackage{textcomp}

\newcommand{\vect}[1]{\mathbf{#1}}

\newcommand{\arctanh}[1]{\operatorname{arctan}}

\begin{document}


\title{\textit{Ab initio} study on the magneto-structural properties of MnAs}


\author{Ivan Rungger}
\author{Stefano Sanvito}
\affiliation{School of Physics, Trinity College, Dublin 2, Ireland}


\date{\today}

\begin{abstract}

The magnetic and structural properties of MnAs are studied with
\textit{ab initio} methods, and by mapping total energies onto a
Heisenberg model.  The stability of the different phases is found
to depend mainly on the volume and on the amount of magnetic
order, confirming previous experimental findings and
phenomenological models.  It is generally found that for large
lattice constants the ferromagnetic state is favored, whereas for
small lattice constants different antiferromagnetic states can be
stabilized. In the ferromagnetic state the structure with minimal
energy is always hexagonal, whereas it becomes orthorhombically
distorted if there is an antiferromagnetic component in the
hexagonal plane.  For the paramagnetic state the stable cell is
found to be orthorhombic up to a critical lattice constant of
about 3.7 \AA, above which it remains hexagonal. This leads
to the second order structural phase transition between
paramagnetic states at about 400 K, where the lattice parameter
increases above this critical value with rising temperature due to
the thermal expansion. For the paramagnetic state an analytic
approximation for the magnitude of the orthorhombic distortion as
a function of the lattice constant is given. Within the mean
field approximation the dependence of the Curie temperature on the
volume and on the orthorhombic distortion is calculated. For
orthorhombically distorted cells the Curie temperature is much
smaller than for hexagonal cells. This is mainly due to the fact
that some of the exchange coupling constants in the hexagonal
plane become negative for distorted cells. This is also the reason
for the appearance of canted spin structures at low temperatures
and high pressures, where the cell is orthorhombic.  With these
results a description of the susceptibility as function of
temperature is given, where the temperature dependence enters via
the dependence of the Curie temperature on the lattice parameters.

\end{abstract}


\maketitle

\section{\label{sec:intro}Introduction}

MnAs is an extremely promising material for magneto-electronics,
since it can grow epitaxially on GaAs \cite{ploog1} and Si
\cite{akeura1} forming clean and atomically sharp interfaces. \cite{akeura1}
MnAs/GaAs heterojunctions have been extensively studied
experimentally, \cite{ploog1,ploog2,song1,song2} and spin
injection from MnAs into GaAs has been demonstrated.\cite{daeweritz1}
However one of the major drawbacks for the use of MnAs in devices is the
fact that bulk MnAs has a phase transition at 318~K, where the
magnetic state changes from ferromagnetic to paramagnetic.
Moreover when grown on GaAs, this temperature changes depending on
the growth direction. This is mainly attributed to the induced
strain.  \cite{ploog1,ploog2,iikawa1,mnas1} The aim of this paper
is to use \textit{ab initio} density functional theory (DFT) calculations
to develop a theory of the phase transitions of MnAs, which can be 
compared with experiments and with existing phenomenological models.

First a review of the experimental properties of MnAs is
presented, and a brief description of the existing
phenomenological models is given. Then the results
of our \textit{ab initio} calculations are presented and compared
to experiments and phenomenological models. Within the scope of a 
Heisenberg model the exchange coupling constants are calculated for different
distorted unit cells, and the Curie temperature and its dependence on the lattice
parameters are evaluated in the mean field approximation. 
The use of the Heisenberg model also makes possible
the prediction of the ground state volume and lattice structure
for the paramagnetic state. It will be shown that the phase
transitions of MnAs can indeed be explained by \textit{ab initio}
calculations. In the last section our results are summarized
and a simple model for the susceptibility as function of
temperature is given. From these a semi-quantitative description of the phase
diagram of MnAs will emerge. 

\section{Experimental properties and existing models}

\subsection{Experimental properties}
\label{sec:expprop}

\begin{figure}[tb] 
\center
\includegraphics[width=8cm,clip=true]{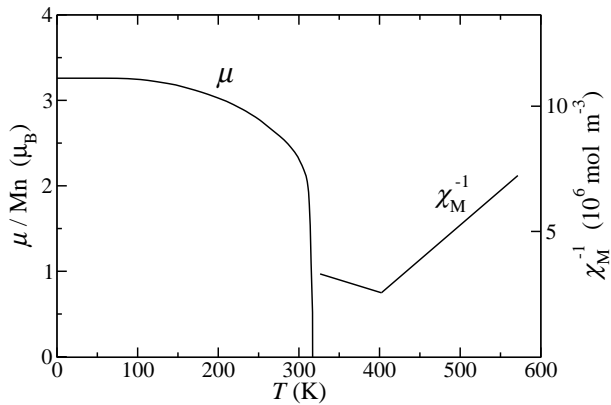}
\caption{\label{fig:mofT}Magnetization per Mn atom $\mu$ as a function
of temperature for ferromagnetic MnAs below 318 K, and inverse
susceptibility $\chi_\mathrm{M}^{-1}$ for paramagnetic MnAs above 318 K (schematically after reference
[\onlinecite{goodek2}]).}
\end{figure}

MnAs is a ferromagnetic metal at low temperature, but it becomes
paramagnetic at $T_\mathrm{p}=318$ K. At this critical temperature
the magnetic moment changes abruptly from a finite value to zero
(figure~\ref{fig:mofT}), the resistivity increases
discontinuously, \cite{mira1} the volume is reduced by 2.1\%, and
the lattice structure changes from the hexagonal B$8_1$ (NiAs-type) to
the orthorhombic B31 (MnP-type) structure.
\cite{suzuki1,beanrodb2,debrod1,goodek1,goodek2} The B31 structure
is obtained by slightly distorting the B$8_1$ structure.
Associated with this transition there is a latent heat of 7490
J/kg. \cite{beanrodb2}  Furthermore the phase transition shows
hysteresis with a critical temperature of 307~K upon
cooling and of 318~K upon heating.\cite{goodek1}
All these properties clearly indicate that the phase transition is of first
order.

Above $T_\mathrm{p}$ the distortion of the crystal structure
reduces continuously, until it vanishes at
$T_\mathrm{t}\approx398$ K. \cite{suzuki1} At $T_\mathrm{t}$ MnAs
undergoes another phase transition, where it changes back from the
B31 structure to the B$8_1$ structure. The lattice dimensions
change continuously and there is no latent heat, but only a
discontinuity in the heat capacity of the material. Therefore this
phase transition is of second order.  For temperatures between
$T_\mathrm{p}$ and $T_\mathrm{t}$ the paramagnetic susceptibility
of the material has an anomalous behavior. It increases with
increasing temperature until it reaches a maximum at
$T_\mathrm{t}$. Above $T_\mathrm{t}$ it decreases and has a
Curie-Weiss behavior (figure~\ref{fig:mofT}).  Moreover at
$T_\mathrm{t}$ there is a lambda point in the specific heat.
\cite{gronvold1}  Application of a magnetic field is found to
transform the B31 structure back to the B$8_1$ above a critical
field. \cite{zieba,zieba2,chernenko1,ishikawa1,mira1}  Table
\ref{tab:propmnaszp} gives a short summary of the described
properties of MnAs.

\begin{table*}[thb] 
\center
\begin{tabular}{*{4}{l}}
\hline\hline\\
                      &  $0 < T < T_\mathrm{p}$ & $T_\mathrm{p} < T < T_\mathrm{t}$ &  $T_\mathrm{t} < T$           \\
\hline\\
 Crystal structure    & Hexagonal B$8_1$        & Orthorhombic B31                  & Hexagonal B$8_1$              \\
 Magnetic order       & Ferromagnetic           & Paramagnetic                      & Paramagnetic                  \\
 Magnetic moment      & $3.4 \mu_\mathrm{B}$    & $--$                              & $--$                          \\
 Susceptibility $\chi$& $--$                    & $\partial\chi/\partial T > 0$     & $\partial\chi/\partial T < 0$ \\
\hline\hline
 \end{tabular}
 \caption{\label{tab:propmnaszp}Some properties of MnAs at zero
pressure.\cite{goodek1}}
 \end{table*}

\begin{figure}
\center
\includegraphics[width=8.5cm,clip=true]{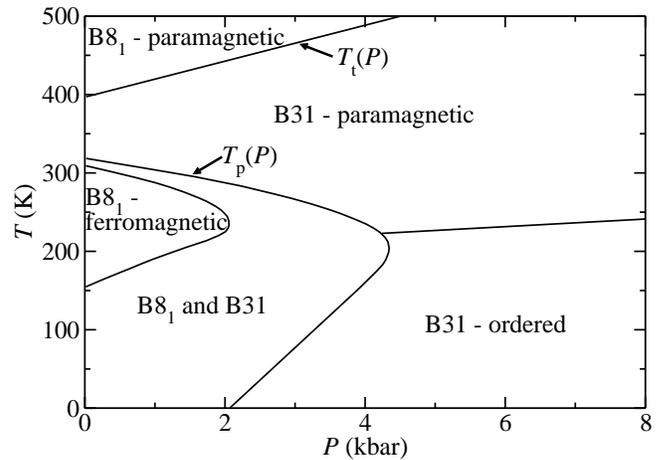}
\caption{\label{fig:phased}Temperature ($T$) versus pressure ($P$)
phase diagram of MnAs (adapted from reference
[\onlinecite{goodek2}]), indicating also $T_\mathrm{p}$ and
$T_\mathrm{t}$ as functions of pressure.}
\end{figure}

Figure \ref{fig:phased} shows the phase diagram of MnAs. If
pressure is applied to the system, $T_\mathrm{p}$ is lowered,
while $T_\mathrm{t}$ increases. Above a critical pressure of 4.6
kbar the ferromagnetic B$8_1$ structure becomes unstable, and the
material has the B31 structure for all temperatures below $T_\mathrm{t}$. At high
pressures and low temperatures different types of ordered magnetic
structures are found, with a reduced saturation magnetic moment
with respect to the zero pressure ferromagnetic phase. This,
together with the anomalous behavior of the susceptibility between
$T_\mathrm{p}$ and $T_\mathrm{t}$, led Goodenough to the idea that
the magnetic moment of the Mn atoms changes from a high spin state
in the B$8_1$ structure to a low spin state in the B31 structure.
\cite{goodek2,goodek1} According to his idea in the B31
structure the moment of the Mn atoms lies between 1 and 2
$\mu_\mathrm{B}$.  However neutron scattering experiments on MnAs
samples in the B31 structure have shown that the local magnetic
moment of the Mn atoms lies between 3 and 3.4 $\mu_\mathrm{B}$ for
different temperatures and pressures, \cite{schwartz1} in clear
contradiction to the assumption of a low spin state. Canted spin
structures, similar to the helimagnetic structures of MnP,
\cite{forsyth1} are found at a pressure of 4.75 kbar below 210 K,
with a local magnetic moment of about 3 $\mu_\mathrm{B}$. The
different saturation magnetic moments for different pressures
therefore correspond to different types of canted magnetic
alignments.  A hysteresis region lies between the ferromagnetic
and the canted regions, where both the B8$_1$ and the B31
structures can be stabilized.

When the magnetic order breaks down and the system becomes
paramagnetic, it has the B31 structure for all pressures. As the
temperature is further increased the structure of the cell
continuously changes back towards the B$8_1$, until at
$T_\mathrm{t}$ it has again the B$8_1$ structure, with $\partial
T_\mathrm{t}/\partial P > 0$, where $P$ is the pressure. 

The magnetocrystalline anisotropy is quite strong in MnAs, with
the $c$-axis being the hard axis, so that the moments prefer to
lie in the hexagonal plane. \cite{debrod1}  Measurements on the
magnetoelastic coupling \cite{ploog3} indicate that the coupling
is stronger in the hexagonal plane than perpendicular to it.

\begin{figure}[ht]
\centering
\subfigure[]{
	\label{fig:uc:hex}
	\includegraphics[width=7.5cm,clip=true]{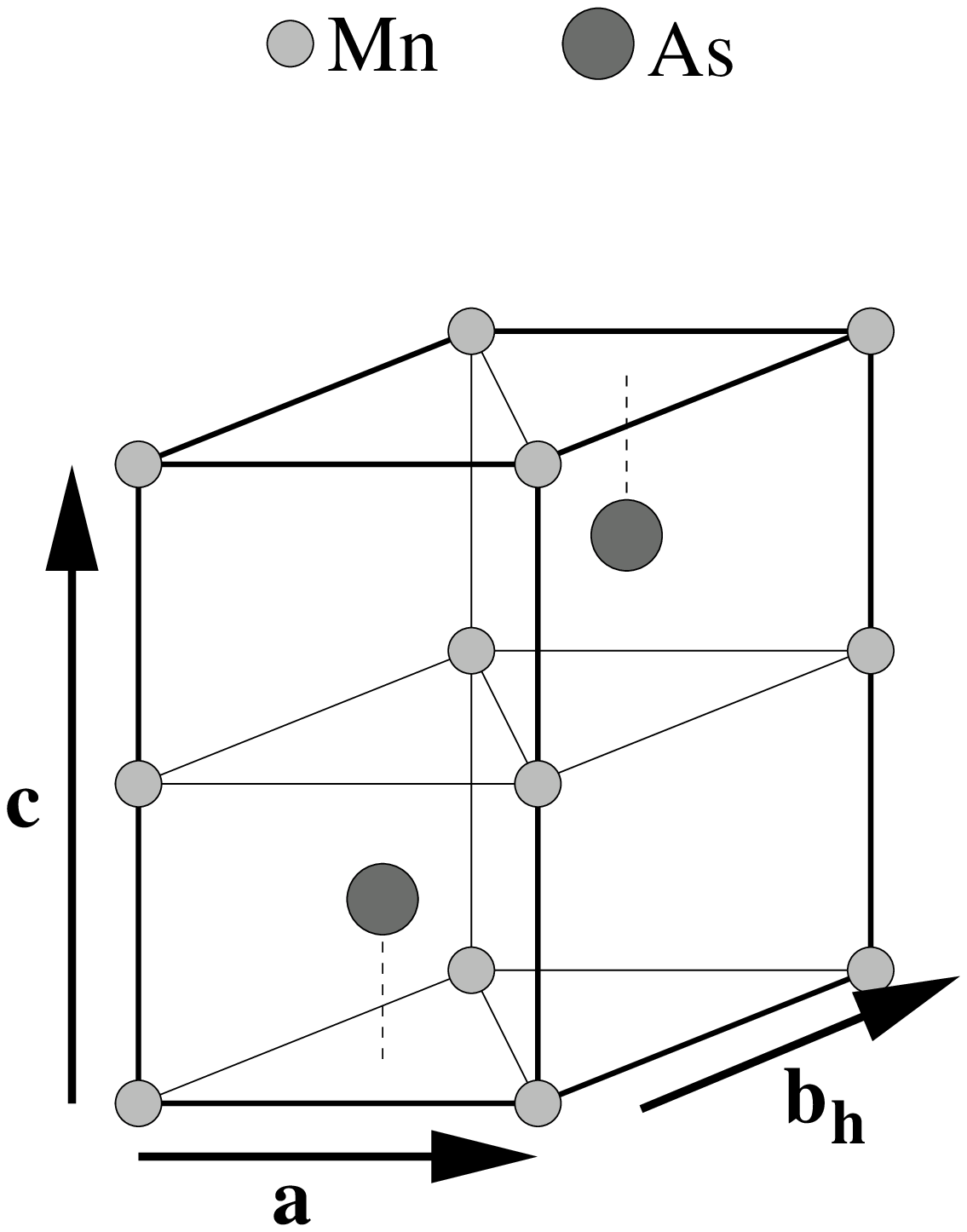}}
\subfigure[]{
	\label{fig:uc:ortho}
	\includegraphics[width=7.5cm,clip=true]{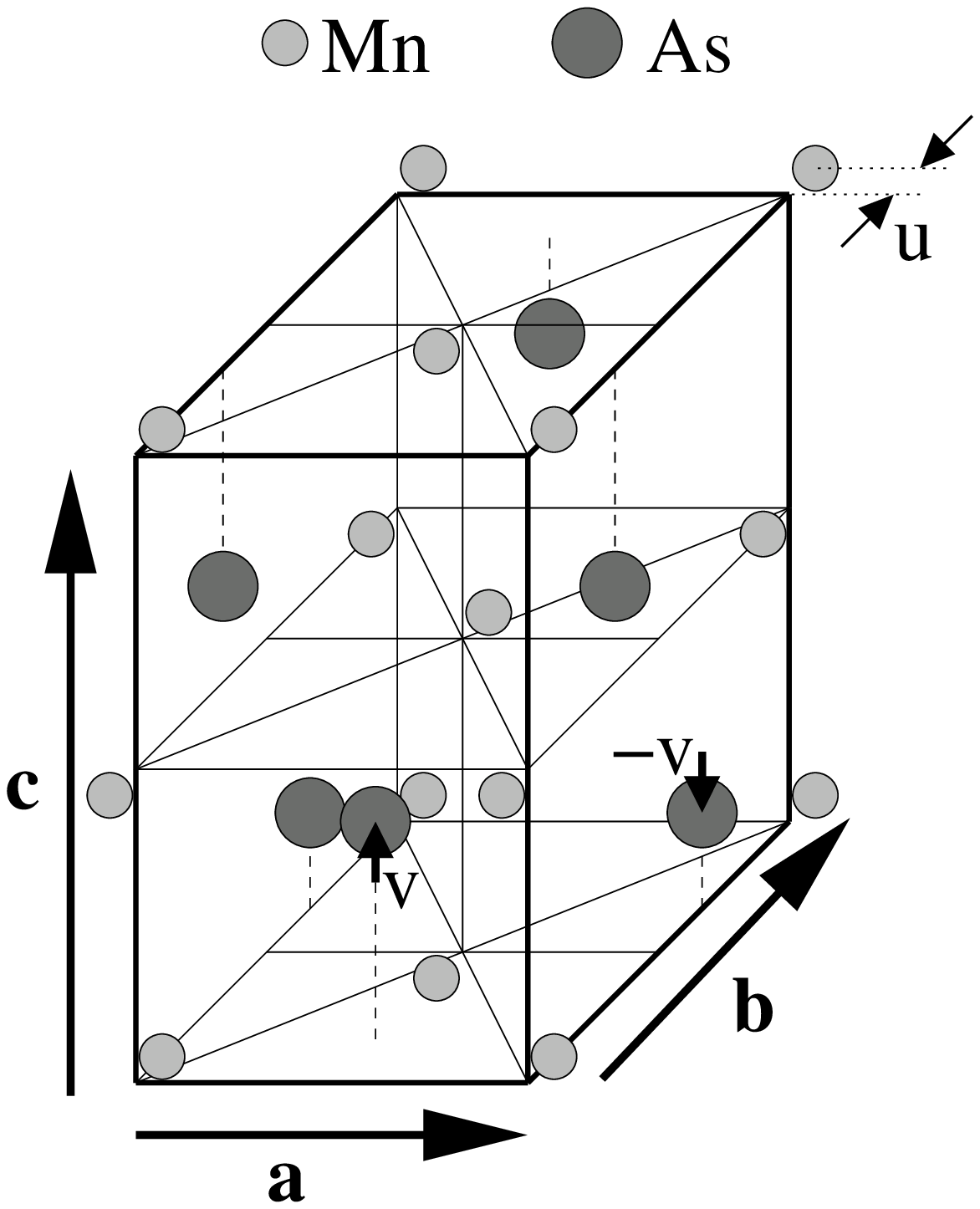}}
\caption{(a) B$8_1$ unit cell containing two Mn and two As atoms, (b)
B31 unit cell containing four Mn and four As atoms.}
\label{fig:uc}
\end{figure}

Figure \ref{fig:uc} shows the unit cells of MnAs in the hexagonal
B$8_1$ and in the orthorhombic B31 crystal structures and it 
defines the unit cell vectors $\vect{a}$, $\vect{b}$ ($\vect{b}_\mathrm{h}$ for 
the B$8_1$ structure) and $\vect{c}$. The B$8_1$
structure consists of stacked hexagonal layers of Mn and As atoms, and
the unit cell contains two Mn and two As atoms. The B31 structure
has twice the volume of the B$8_1$ due to symmetry lowering
and contains four Mn and four As atoms. The lattice is
nearly hexagonal, and the atoms are moved out of the hexagonal
symmetry points along the $\vect{b}$ and $\vect{c}$ directions
(figure \ref{fig:uc:ortho}). 

\begin{figure}[ht]
\centering
\includegraphics[width=6.5cm,clip=true]{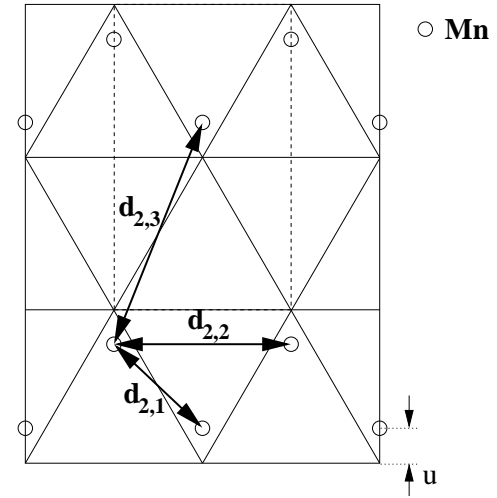}
\caption{Two-dimensional representation of one layer of Mn atoms in
the B31 structure. d$_{ij}$ represent the various Mn-Mn distances. The first
index $i$=2 indicates that all the Mn are second nearest-neighbor 
to each other in the B8$_1$ structure. The second index $j=$1, 2, 3 labels 
the three distances arising from the B31 distortion.
}
\label{fig:ortho_2d}
\end{figure}

The Mn atoms are mainly displaced in
the hexagonal plane along the $\vect{b}$ direction, to form chains
of closer Mn atoms, separated from the other chains by a larger
distance (figure \ref{fig:ortho_2d}). The As atoms are mainly
displaced along the $\vect{c}$ axis. In each unit cell one of the planar
As atoms is moved upwards and the other downwards with
respect to the original position in the B8$_1$ structure, so as to
keep the Mn-As distance nearly constant.  The
displacement $u$ of the Mn atoms in the hexagonal plane lies
between 0 and $0.05~b$ ($b=|\vect{b}|$), depending on the
temperature and pressure, and the displacement $v$ of the As atoms
along the $c$-axis lies between 0 and $0.05~c$ ($c=|\vect{c}|$).
The B$8_1$ structure is a special case of the B31 structure, where
$b=\sqrt{3}~a$ and $u=v=0$. 

Therefore we choose the unit cell vectors in such a way that
$\vect{a}$ and $\vect{c}$ have the same direction for both the B8$_1$
and the B31 structures. In contrast the directions of the vectors
$\vect{b}_\mathrm{h}$ ($|\vect{b}_\mathrm{h}|=|\vect{a}|=a$) for
the hexagonal cell and $\vect{b}$ for the orthorhombic cell are
different.

\begin{figure*} 
\center
\includegraphics[width=15cm,clip=true]{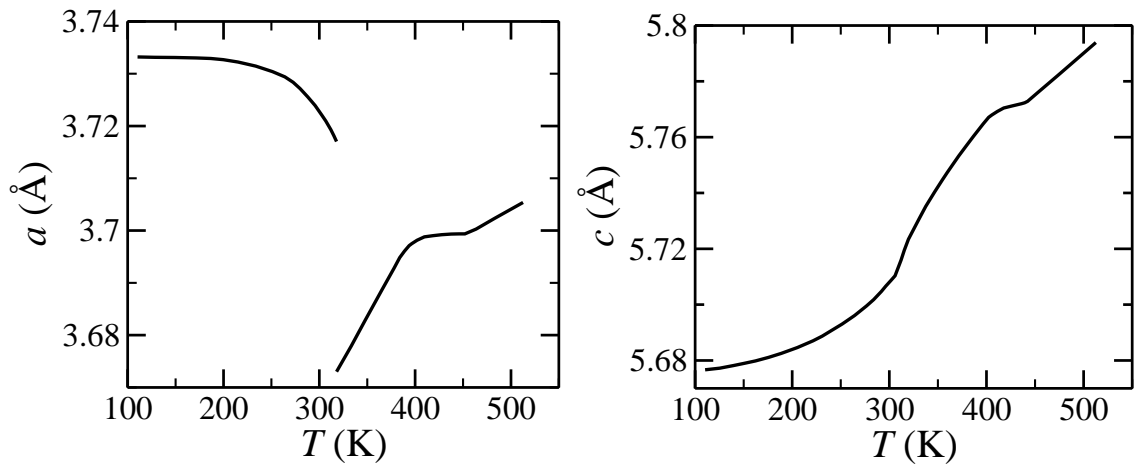}
\caption{\label{fig:acofT}Lattice parameters $a$ and $c$ (see figure
\ref{fig:uc}) as a function of temperature (after
reference [\onlinecite{suzuki1}]).} 
\end{figure*}

The lattice parameters of MnAs as function of temperature and
magnetic field have been measured in several works
[\onlinecite{willis1,suzuki1,wilsonk,chernenko1}]. Figure
\ref{fig:acofT}
shows the change of the lattice parameters as a function of
temperature for zero pressure.\cite{suzuki1} The lattice parameters increase
with temperature due to normal thermal expansion.  However the in
plane lattice parameter $a$ decreases when the temperatures get
near $T_\mathrm{p}$, where it jumps from $3.717$ \AA~to the lower
value of $3.673$ \AA. The perpendicular lattice parameter $c$
always increases continuously with temperature. At $T_\mathrm{t}$
there is an inflection in the slope of the lattice parameter as a
function of temperature, and at about $T_\mathrm{s}=450$ K the
slope changes discontinuously. In other measurements
the discontinuous change of the slope is found at
about 410 K.\cite{wilsonk}

The exact temperature at which the distortion
disappears is somewhat uncertain, and fluctuations may play a role
for small distortions. The given temperature for the disappearing
of the distortion correspond to $T_\mathrm{t}=398$ K,
\cite{suzuki1,wilsonk} however measurements for small distortions
are difficult and such temperature can only be inferred. As
pointed out in reference [\onlinecite{zieba1}] the distortion
should appear at temperatures slightly above $T_\mathrm{t}$. Throughout
this work we assume that the disappearing of the
distortion occurs at $T_\mathrm{s}$, which the temperature where the thermal
expansion coefficient of MnAs changes abruptly.

\subsection{\label{sec:revie}Review of existing models}

Until the fifties it was believed that the first order phase transition at $T_\mathrm{p}$ 
was between a ferromagnetic and an antiferromagnetic state. Kittel then proposed a
model where the distortion from the B$8_1$ to the B31 structure produces the
change in sign of one of the exchange coupling constants, giving rise to 
antiferromagnetic order.\cite{kittel_exinv} However there was no experimental evidence of
this antiferromagnetic state and experiments demonstrated that the
transition is instead to a paramagnetic state. \cite{wilsonk} 

Therefore Bean and Rodbell (BR) proposed a modification of Kittel's theory,
where the exchange interactions are ferromagnetic for both
structures, but they are much weaker in the B31 phase.  \cite{beanrodb1,beanrodb2} 
The main idea is that the exchange interaction decays strongly as the volume 
decreases. In order to simplify the model, the analysis was based 
on the extrapolated Curie temperature $T_\mathrm{C}$ only, and not on the
details of the magnetic interaction at the atomic level.
Furthermore the model neglected the anisotropic change of the crystal structure, 
and assumed that it is possible to describe the change from the B$8_1$ to the B31
structure by a change in volume only. The dependence of $T_\mathrm{C}$ on the 
volume $V$ was described by the following equation
\begin{equation}
T_\mathrm{C}=T_\mathrm{0}\left[1+\beta~\frac{\left(V-V_0\right)}{V_0}\right]~,
\label{eq:tcbr}
\end{equation} 
where $T_0$ is the Curie temperature at the volume $V_0$, which is the volume 
that the system would have in the hypothetical absence of exchange interaction. 
In this context this corresponds to the volume of the B31 structure above
$T_\mathrm{p}$. $\beta$ is a parameter and is determined by
fitting the model to experiments. Within the BR model one can show
that for certain values of $\beta$ a first order phase transition
between a ferromagnetic and paramagnetic state occurs with a
simultaneous change of the volume. However the model does not explain 
the second order phase transition at $T_\mathrm{t}$. Further improvements such as
the introduction of a term proportional to $(V-V_0)^2$ do not change 
the main results.\cite{baerner1} 

Later Goodenough made an attempt to explain the anomalous behavior
of the susceptibility and the second order phase transition at
$T_\mathrm{t}$ by extending the BR model and assuming that the
local magnetic moment on the Mn atoms depends strongly on the
volume. \cite{goodek2,goodek1} Here the Mn atoms are assumed to be
in a high spin state in the B$8_1$ structure and in a low spin
state in the B31 structure. \cite{goodek2,goodek1} However
measurements of the local magnetic moment show that the change in
the magnetic moment at $T_\mathrm{p}$ is very small, therefore the
Goodenough model is not applicable to this phase transition.  This
is probably due to the fact that the MnAs unit cell volume in any
crystalline structure is too big to justify a high-spin to low-spin transition. In
fact in MnAs$_{1-x}$P$_x$ the change from high-spin to low-spin
state is observed, \cite{suzuki1,haneda1,haneda2} however the low
spin state is found for unit cell volumes smaller than 120 \AA$^3$. In
contrast the unit cell of MnAs has always a volume of around 130
\AA$^3$, and it is always in the high-spin state.

In 1982 Kato et al. extended the BR model by taking into account not
only the change in volume, but also the change of the crystal
structure. \cite{katonagai1} Furthermore instead of just using
$T_\mathrm{C}$ for describing the magnitude of the exchange
interactions, they consider exchange coupling constants up to second
nearest neighbors. The obtained results are similar to those of Bean and Rodbell.
The second order phase transition at $T_\mathrm{t}$ is explained
by assuming that locally the structure above $T_\mathrm{t}$ is
still the B31, but that the distortions from the B$8_1$
are randomly distributed. This assumption is probably not valid, since neutron
diffraction experiments clearly indicate that the structure is a
regular B$8_1$ above $T_\mathrm{t}$. \cite{wilsonk}

A Landau-type phenomenological model, where the free energy $\Phi$
is expanded as a function of order parameters, is given in
references [\onlinecite{zieba1,kamenev1,kamenev2}] and references
therein. In reference [\onlinecite{zieba1}] just two order
parameters are used, the relative magnetization $\sigma$
($\sigma=1$ for a ferromagnetic state, $\sigma=0$ for a
paramagnetic state) and the orthorhombic distortion $d$. The
following equation is used for the free energy:
\begin{equation}
\begin{split}
\Phi(d,\sigma;&T,H)=\Phi_0 + c_1 \left(T-T_0\left(1- \delta_1~ d^2\right)\right)
\sigma^2 + c_2~\sigma^4 + \\
+& c_3 \sigma^6 +c_4 \left(T-T_\mathrm{D}\right)d^2+ c_5 d^4 - M_0
\sigma H \left(1- \delta_2 d^2\right),
\end{split}
\label{eq:ziebaeq}
\end{equation}
where $c_1,...,c_5$ and $\delta_1,\delta_2$ are expansion
coefficients to be fitted to experiment, $H$ is an external
magnetic field, $M_0$ is the saturation magnetization, $T_0$ is
the extrapolated Curie temperate of the low temperature phase, and
$\Phi_0$ is a constant. 

The distortion $d$ plays the same role as
the relative change in volume $(V-V_0)/V_0$ of the BR model
(equation \eqref{eq:tcbr}). In equation \eqref{eq:tcbr} $T_\mathrm{C}$ depends
linearly on the change in volume, whereas now it depends
quadratically on the distortion $d$, $T_\mathrm{C}=T_0(1-
\delta_1~ d^2)$. This is the correct expansion of $T_\mathrm{C}$,
since the linear term in $d$ disappears due to symmetry ($+d$ and
$-d$ correspond to the same distortion). The terms in $\sigma^2$, $\sigma^4$ and $\sigma^6$ appear
also in the BR model and lead to the first order transition at
$T_\mathrm{p}$. The second order phase
transition at $T_\mathrm{t}=T_\mathrm{D}$ is generated by the
$d^2$ and $d^4$ terms in the expansion. 

Also the variation of the
magnetic moment with the distortion is contained in the model,
although the authors find that the coefficient $\delta_2$ is
essentially zero. For $T_0$ the Curie temperature extrapolated from the high 
temperature susceptibility above $T_\mathrm{t}$ was considered 
($T_0=285$~K), and for $T_\mathrm{D}$ the temperature at which the inverse 
susceptibility has a minimum ($T_\mathrm{D}=394$~K). By construction this model
yields the correct thermodynamic behavior, and also predicts the
increase of the susceptibility between $T_\mathrm{p}$ and
$T_\mathrm{t}$ by means of the reduction of the distortion with
increasing temperature. 

Variations of this model
[\onlinecite{kamenev1,kamenev2}] give similar results. In
reference [\onlinecite{kamenev2}] the full $T$-$P$-$H$ phase diagram
of MnAs is explained with a Landau-type expansion, where more
order parameters are used. However also in this case a term
equivalent to $(T-T_\mathrm{D})d^2$ of equation \eqref{eq:ziebaeq}
is used in order to obtain the phase transition at $T_\mathrm{D}$.
Despite the fact that Landau-type expansions give very good
agreement between theory and experiments when the right parameters
are used, they do not provide insights into the origin of the terms
of the expansion, especially of the $(T-T_\mathrm{D})d^2$
term. In reference [\onlinecite{valkov2}] a basic justification of
such term is given from first principles within a spin fluctuation
theory constructed from a Hubbard Hamiltonian. It is shown that
for a given volume the minimum of the free energy can lie at $d=0$
for a ferromagnetic state or at $d\ne0$ for a paramagnetic state.

In 1986-87 Motizuki and Katoh used spin fluctuation theory in
order to explain the anomalous behavior of the susceptibility
between $T_\mathrm{p}$ and $T_\mathrm{t}$. \cite{katoh1,motizuki2}
A Hubbard Hamiltonian was used, with model density of states
obtained from first principle calculations. They could
qualitatively show that the susceptibility increases when going
from the B31 structure to the B$8_1$ structure, again mainly due
to the fact that $T_\mathrm{C}$ increases with increasing
temperature. 

More recently various tight binding \cite{pod1,pod2} as well
as first principles\cite{stefano1,ravindran1,freeman1,freeman2,paiva1,
debernardi1,debernardi2,shirai1,motizukik1} calculations have been
performed for MnAs in the B$8_1$ structure. The results generally agree
and compare well with the experiments. Only two studies on MnAs in
the B31 structure are known to the authors.
\cite{kleinman1,valkov1} In reference [\onlinecite{kleinman1}] a
description of the paramagnetic state of the B31 structure is
given by assuming that the paramagnetic state coincides with zero
local magnetic moment of the Mn atoms. This does not correspond to
the usual way of describing paramagnetism, which rather
corresponds to constant local magnetic moments, whose directions
in space change randomly in time due to spin fluctuations.

The present work investigates the magnetic interactions across the
various phase transitions of MnAs. An explanation of the magnetostructural properties
in terms of first-principles calculations is given, thereby illustrating the origin and providing a
justification of the parameters used by the different models.

\section{Results}
\subsection{\label{sec:dft}DFT calculations}

First principles calculations within density functional theory
(DFT) are performed using the pseudopotential code based on localized atomic
orbitals SIESTA. \cite{siesta} The generalized gradient
approximation (GGA) is used for the exchange correlation
potential,\cite{pbe0}  since it has been shown to give good structural
properties for hexagonal MnAs. \cite{stefano1,freeman1,freeman2}
In the valence we consider $4s, 4p$ and $3d$ orbitals for Mn, and
$4s$, $4p$ and $4d$ for As.  For both Mn and As double zeta
polarized local orbitals are used for the $s$ and $p$ angular
momenta, whereas for the $d$ orbitals double zeta
is used. The number of $k$-points in the Brillouin zone is
specified by a grid cutoff of 20 \AA. This corresponds to a
11$\times$11$\times$8 mesh for the B$8_1$ unit cell, giving approximately 
1000 $k$-points in the full Brillouin zone. For the B31 unit cell such cutoff 
yields a 8$\times$11$\times$7 mesh.  The real space mesh cutoff, which 
determines the density of the real space grid, is 300 Ry.

After full relaxation of the unit cell to a pressure below 0.1
kbar, and of the atomic positions to a force smaller than 0.01
eV/\AA~the B31 unit cell in the ferromagnetic configuration
relaxes to a B$8_1$ structure with $a=3.72$ \AA ~and $c=5.58$ \AA.  
The experimental values at room
temperature are $a=3.724$ \AA ~and $c=5.707$ \AA. Therefore the
relative error is below $1\%$ for $a$ and $-2$\% for $c$. The
lattice parameters at 110 K can be extracted from figure
\ref{fig:acofT} and are $a=3.733$ \AA~and $c=5.677$ \AA. This
demonstrates that GGA-DFT reproduces rather well the zero
temperature ground state.

For a fixed $c/a$ ratio of 1.54, which is close to the
experimental value at the first order phase transition, the energy is minimized 
for $a=3.695$ \AA, which compares well with the results of other \textit{ab initio}
calculations. \cite{stefano1,freeman1,freeman2}  Also the band
structure and the density of states are similar to previous
calculations.  The magnetic moment per Mn atom is
$3.4~\mu_\textrm{B}$, and compares well with the measured value of
$3.4~\mu_\textrm{B}$.\cite{goodek1}
\begin{figure} \centering
\includegraphics[width=7.5cm,clip=true]{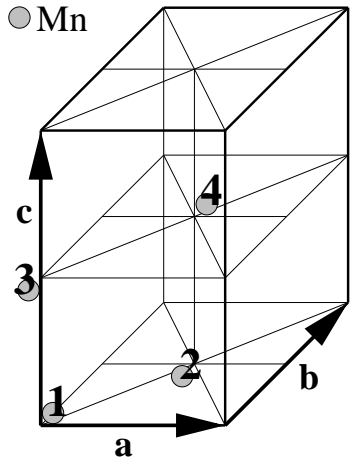}
\caption{B31 unit cell where only the four Mn atoms are shown displaced
out of the B8$_1$ symmetry positions.}
\label{fig:uc_ortho_numMn}
\end{figure}

The unit cell of the B31 structure contains 4 Mn atoms, allowing
for 3 possible independent antiferromagnetic configurations of the
local moments of the Mn atoms. The different antiferromagnetic
states are $++--$, $+-+-$ and $+--+$.  As a matter of notation
$++--$ means that the atoms 1 and 2 in the unit cell have opposite
magnetic moment than that of atoms 3 and 4. The indices of the Mn
atoms in the unit cell are defined in figure
\ref{fig:uc_ortho_numMn}.  A cell relaxation is performed for
those three antiferromagnetic configurations. Table
\ref{tab:relaxcell} lists the obtained relaxed structures together
with the total energies per Mn atom as compared to the
ferromagnetic ground state energy ($E-E_\mathrm{FM}$). The
structure remains of the B$8_1$ type if the local moments are
ferromagnetically aligned in the hexagonal plane, whereas it
changes to the B31 type if the moments are antiferromagnetically
aligned, with $u$ ($v$) of the order of 0.05 $b$ ($c$). There is
also a slight displacement of the Mn atoms along $\vect{c}$ of at
most 0.01 $c$, and of the As atoms along $\vect{b}$ of at most
0.01 $b$. 

Generally it can be observed that the in-plane lattice parameters
contract and the $c$-parameter expands for the antiferromagnetic
states, resulting in a reduction of the volume $V$. The calculated
values are similar to those given in reference
[\onlinecite{kleinman1}].

The total energy for the ferromagnetic alignment is the lowest,
although the $+--+$ configuration is higher of only 17 meV/Mn.
This indicates that the system should evolve to one of the
antiferromagnetic states under pressure, since those have a much
smaller volume but only a slightly higher energy.

 \begin{table*}[tb] 
\hspace{-0.5in}
 \begin{tabular}{c|*{11}c}
   &~~$a($\AA$)$ & ~~$b($\AA$)$ & ~~$c($\AA$)$ & ~~$V($\AA$^3)$ & ~~~$u/b$~ & ~~$v/c$~ & ~$d_{2,1}$(\AA) & ~$d_{2,3}$(\AA) & $\mu_\mathrm{Mn}(\mu_\textrm{B})$ & $\mu_\mathrm{As}(\mu_\textrm{B})$&  $E-E_\textrm{FM}(\textrm{meV})$\\ \hline\\
 $++++$  ~ &~  3.72 & 6.47 & 5.58 & 134.27 & 0.00 & 0.00 & 3.73 & 3.73 & 3.43 & -0.24&  0 \\
 $++--$  ~ &~  3.56 & 6.18 & 5.81 & 127.93 & 0.00 & 0.00 & 3.56 & 3.57 & 3.10 & ~0.00& 62 \\
 $+-+-$  ~ &~  3.55 & 6.24 & 5.62 & 124.54 & 0.05 & 0.05 & 3.10 & 4.10 & 3.01 & -0.08& 17 \\
 $+--+$  ~ &~  3.62 & 6.29 & 5.70 & 129.83 & 0.04 & 0.04 & 3.12 & 4.17 & 3.33 & -0.03& 35 \\
 \end{tabular}
 \caption{\label{tab:relaxcell}Relaxed lattice parameters, local
magnetic moment of the Mn and As atoms, and total energies per Mn atom for
different spin configurations.}
 \end{table*}

Table \ref{tab:relaxcell} gives also the distances between
the first three nearest neighbor Mn atoms in the
hexagonal plane $d_{2,1}$, $d_{2,2}$ and $d_{2,3}$ (see figure
\ref{fig:ortho_2d}). Note that $d_{2,2}=a$ and it is not given
explicitly. While these distances are all equal in the hexagonal
case, they differ of as much as 1 \AA ~in the B31 structure.
Large changes in the distance between the Mn atoms are
possible since the nearest neighbor Mn-Mn separation in MnAs is
well above the inter-atomic distance 2.61 \AA~of bulk Mn,
\cite{asada} which can be regarded as the minimal possible
distance between Mn atoms. The distance between nearest neighbor
Mn and As atoms lies between 2.46 \AA ~and 2.62 \AA~for all the
different configurations, and changes therefore much less than the
Mn-Mn distance.

The local magnetic moment on the Mn ($\mu_\mathrm{Mn}$) and As
($\mu_\mathrm{As}$) atoms, calculated using the atomic Mulliken
population, \cite{mulliken0} is also given in table \ref{tab:relaxcell}. The local
moment on the Mn atoms ranges between 3.43 $\mu_\mathrm{B}$ for
the ferromagnetic configuration to 3.01 $\mu_\mathrm{B}$ for the
$+-+-$ configuration. This reduction in the local moment is mainly
due to the decrease of the cell volume, and the consequent
increase of the hybridization between the Mn-$d$ and As-$p$
orbitals.

In summary these calculations show that the distortion to the B31
structure is caused by an antiferromagnetic alignment of the local
magnetic moments in the hexagonal plane.

\subsection{\label{sec:heisenberg}Fit to Heisenberg energy}

In order to extract the various exchange parameters, calculations
are performed for three different B31 supercells in different
local magnetic configurations. These supercells contain 8 Mn
atoms, and are obtained by doubling the B31 unit cell along the
$\vect{a}$ lattice vector (supercell 1), along $\vect{b}$
(supercell 2) and along $\vect{c}$ (supercell 3). The calculated
total energies are then fitted to a model Hamiltonian. The most
general form of Hamiltonian able to fit all possible energies for
a system of $N_s$  collinear magnetic moments is
\begin{equation} E_{s_1,s_2,\dots} =
E_0- \sum_{\nu=1}^{N_s} ~\frac{1}{\nu!}~
\sum_{j_1}s_{j_1}\sum_{j_2}s_{j_2}\dots\sum_{j_\nu}
s_{j_\nu}~J_{j_1,j_2,\dots,j_\nu}~,
\label{eq:fit0}
\end{equation}
where $s_j$ is the moment at site $j$ (normalized to one), and the
$J$s are coupling parameters. The $J_{ij,\dots}$ parameters are
symmetric under permutations of the indices, and each of the
$J_{ij,\dots}$ is zero if any of the indices $i$ and $j$ are
equal. The number of independent parameters is therefore
$2^{N_s}$. Since time reversal symmetry makes the
system invariant under a global spin rotation only $J$s with an
even number of indices are not zero and equation \eqref{eq:fit0}
reduces to
\begin{align}
E_{s_1,s_2,\dots} = & E_0 - \frac{1}{2}\sum_{i,j}s_i s_j ~J_{ij}~
- \frac{1}{4!} \sum_{i,j,k,l}s_i s_j s_k s_l ~J_{ijkl}~ - \\ &
- \frac{1}{6!} \sum_{i,j,k,l,m,n}s_i s_j s_k s_l s_m s_n
~J_{ijklmn}~ - \dots \notag
\end{align} 
Here we neglect 4-moment coupling constants $J_{ijkl}$ and
higher thus reducing the model to an Ising-type. This approximation neglects the
dependence of the moment on each Mn atom as well as the small
induced magnetic moment of the As atoms on the orientation of the
moments of the surrounding Mn atoms. Furthermore it is assumed
that the coupling constants are independent from the angle between
the magnetic moments, so that the fit can be extrapolated to
vector magnetic moment $\vect{s}_i$ to give a Heisenberg type
energy
\begin{equation}
E_{\vect{s}_1,\vect{s}_2,\dots} = E_0 -
\frac{1}{2}\sum_{i,j}\vect{s}_j \vect{s}_j ~J_{ij}.
\label{eq:heis} 
\end{equation}
In mean field theory the Curie temperature $T_\mathrm{C}$ for
classical Heisenberg exchanged magnetic moments is
\begin{equation}
k_\mathrm{B} T_\mathrm{C}=\sum_{j}J_{0j}/3=J_0/3,
\label{eq:tc}
\end{equation}
where $k_\mathrm{B}$ is the Boltzmann constant.  The quantity
$J_0=\sum_{j}J_{0,j}$ is the sum of the exchange coupling
constants of a given magnetic moment with all the other moments.
In the following sections equation \eqref{eq:tc} is always used to
extract Curie temperatures, although it is well-known that the
mean field approximation overestimates $T_\mathrm{C}$. \cite{turek_Fe}

Calculations are performed for all the independent spin
configurations of the supercell 1, and for a randomly chosen
subset of those of supercells 2 and 3. 35 different configurations
of the magnetic moments are used in total. The energies are then
fitted by a least-mean-square fit to the coupling parameters of
equation \eqref{eq:heis}. Since the system is metallic with the
$d$-orbitals having finite density of states at the Fermi level,
the magnetic interaction is expected to have a long range
character. For the chosen supercells it is possible to extract
coupling constants up to the ninth nearest neighbor. The lattice
parameters used are approximately those for ferromagnetic MnAs in
the B$8_1$ structure at the phase transition temperature
$T_\mathrm{p}=318$ K ($a$=3.71 \AA, $c/a$=1.54).

\begin{figure} \centering 
\includegraphics[width=8cm,clip=true]{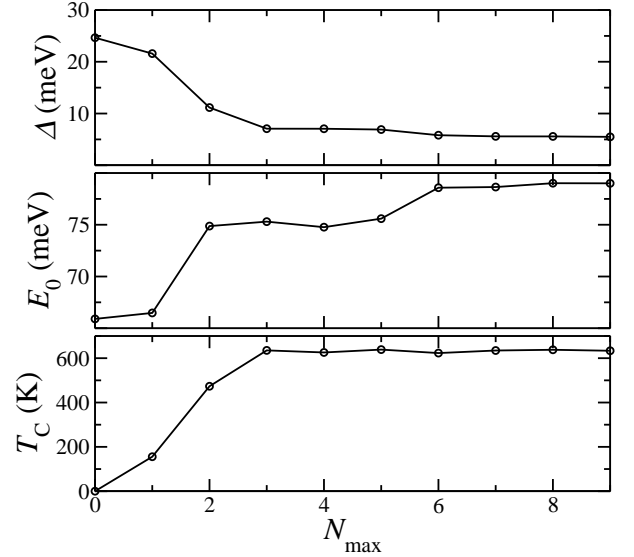} 
\label{fig:numc:tc_nn}
\caption{Variation of the various exchange quantities as function of the number 
of coupling coefficients $N_\mathrm{max}$ included in the fit.
(a) Standard deviation $\Delta$ of the energies resulting
from equation  \eqref{eq:heis} as compared to the calculated DFT energies
per Mn atom. (b) $E_0$ per Mn atom (equation \eqref{eq:heis}), where
the zero of energy is chosen as the energy of the ferromagnetic
state. (c) Mean field Curie temperature $T_\mathrm{C}$.} \label{fig:numc}
\end{figure}

We have carefully tested the convergence of our results with the range of
the Heisenberg exchange interaction. Figure
\ref{fig:numc} shows the standard deviation $\Delta$ of the
energies resulting from equation \eqref{eq:heis} as compared to
the calculated DFT energies per Mn atom, the value of $E_0$ per Mn
atom and the mean field Curie temperature $T_\mathrm{C}$ as a
function of the number of coupling coefficients $N_\mathrm{max}$
included in the fit.
The standard deviation $\Delta$ decays monotonically, remains
roughly constant for $N_\mathrm{max} \ge 3$, and then reaches a
minimum value of around 5 meV for $N_\mathrm{max}=9$.  This can be
considered as the error resulting from neglecting high moment 
coupling constants. The value of $E_0$ changes less over the
whole range. 

$T_\mathrm{C}$ reaches a constant value of approximately 633~K
for $N_\mathrm{max}\ge3$.  This indicates that the main
contribution arises from the first three nearest neighbor coupling constants. 
The experimental value of $T_\mathrm{C}$ for the low temperature phase
ranges between $T_\mathrm{p}=318$ K and $T_\mathrm{t}=400$ K.  This
means that our mean field $T_\mathrm{C}$ overestimates the
experimental one by about a factor 2.
\begin{figure} 
\center
\includegraphics[width=7.0cm]{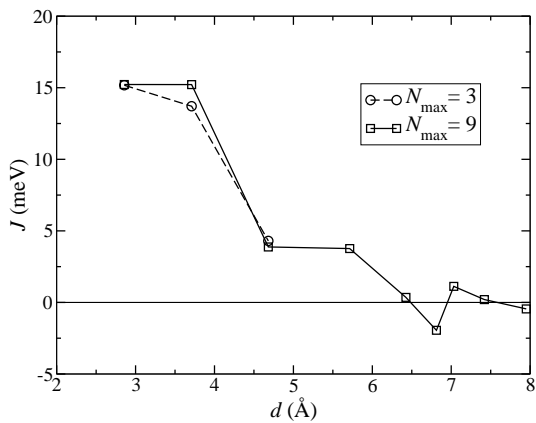}
\caption{\label{fig:j39hex} Exchange coupling parameters for
$N_\mathrm{max}=3$ and $N_\mathrm{max}=9$ as a function of the
distance between the magnetic moments.}
\end{figure}

Figure \ref{fig:j39hex} shows the calculated exchange coupling
constants as a function of the distance for two different fits
counting respectively 3rd and 9th nearest neighbor coupling. The
first three exchange constants remain nearly unchanged when going
from 3rd to 9th nearest neighbor coupling.
Interestingly the coupling parameters are positive and therefore ferromagnetic
up to $d\sim6.5$\AA\ (fifth neighbor interaction). 
In what follows we consider only coupling parameters up to third nearest
neighbors, as they give the main contribution to the
properties of the material.  

\subsection{B8$_1$ to B31 distortion at $T_\mathrm{p}$}
\label{disttp}

\begin{figure}
\centering
\subfigure[]{
	\label{fig:js:hex}
	\includegraphics[width=7.5cm]{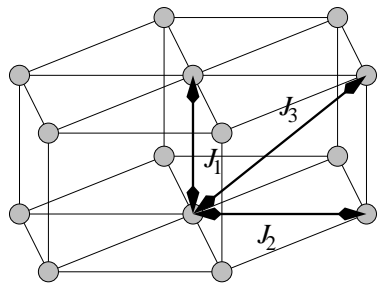}}
\subfigure[]{
	\label{fig:js:ortho}
	\includegraphics[width=7.5cm]{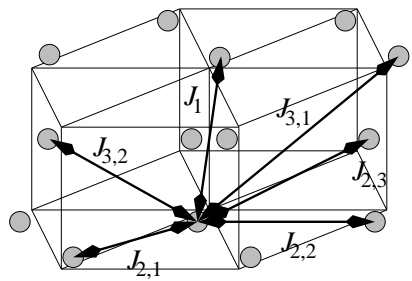}}
\caption{Schematic representation of the atomic positions of the Mn atoms
together with the exchange constants
for the B$8_1$ (a), and the B31 (b) structures.}
\label{fig:js}
\end{figure}

The B$8_1$ to B31 structural phase transition at $T_\mathrm{p}$ is investigated
by calculating the Heisenberg coupling constants for different distorted cells. 
We start from the B$8_1$ with the experimental lattice parameters near
$T_\mathrm{p}$ ($a=3.71$ \AA, $b=\sqrt{3}~a$, $c=1.54~a$, $u=v=0$)
and then distort the cell linearly to the B31 structure. 
The amount of distortion $d$ is given in percent, where $d=$0 \% stands 
for the lattice parameters of the ferromagnetic B$8_1$ cell at $T_\mathrm{p}$, and
$d=100$ \% for the paramagnetic B31 cell at $T_\mathrm{p}$
($a=3.676$ \AA, $b=1.01 \sqrt{3}~a$, $c=1.556~a$, $u=0.02~b$, $v=0.02~c$).
Calculations are done for distortions between 0 \% and 220 \%.
Note that the volume decreases with increasing distortion. For
these calculations only the supercells 1 and 2 are used with a
total of 26 different spin configurations.  The standard deviation
of the fit is approximately constant for all the distortions and
is of the order of 5~meV/Mn.

Figure \ref{fig:js:hex} shows the Mn atoms of the B8$_1$ structure coupled by first ($J_1$), 
second ($J_2$) and third ($J_3$) nearest neighbor interaction.
In the distorted B31 structure the three coupling constants $J_1$, $J_2$ and $J_3$ 
are split into six different constants due to symmetry loss. While there is still only one
$J_1$ coupling, the in-plane $J_2$ splits into three different coupling constants 
$J_{2,1}$, $J_{2,2}$ and $J_{2,3}$, corresponding to different distances between the 
Mn atoms in the hexagonal plane (see figure \ref{fig:ortho_2d}). 
Moreover also the third nearest neighbor coupling $J_3$ splits into three
different constants, although two of them are between Mn atoms
separated by approximately the same distance at $T_\mathrm{p}$, and so they 
are assumed to be identical. Hence $J_3$ effectively splits only into $J_{3,1}$ 
and $J_{3,2}$.

\begin{figure}
\centering
\includegraphics[width=7.5cm,clip=true]{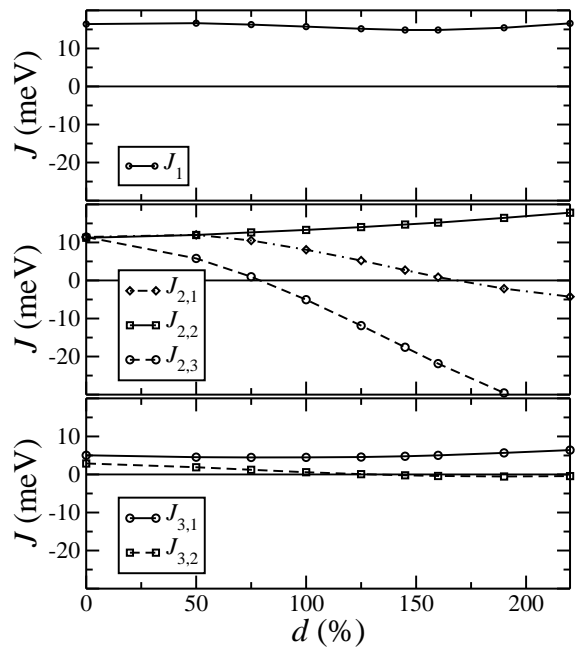}
\caption{Evolution of the exchange coupling constants when distorting the
unit cell linearly from the B$8_1$ structure to the B31 structure.
$d=0$ \% represents the  B$8_1$ structure at $T_\mathrm{p}=318$ K, $d=100$ \%
represents the  B31 structure at $T_\mathrm{p}$. A positive (negative) value of
$J$ means ferromagnetic (antiferromagnetic) coupling.} 
\label{fig:js_od}
\end{figure}

Figure \ref{fig:js_od} shows the calculated values for the
exchange parameters as a function of the distortion. For
0\% distortion the values of $J_{2,1}$, $J_{2,2}$ and $J_{2,3}$
are approximately equal reflecting the hexagonal symmetry. The
values of $J_{3,1}$ and $J_{3,2}$ also should be identical although
they differ by about 2~meV (note that in the fit we do not force the B8$_1$
symmetry when determining the $J$s for the undistorted structure). This can 
be assumed to be the error over the fit. Additional control
fits were also performed for different subsets of the 26 spin
configurations. The variation over the individual $J$s was of 20\%, whereas
the variation of $J_0$ was always smaller than 6\%.  

The value of $J_1$ remains approximately constant for all the distortions, reflecting
the fact that the distance between the quasi-hexagonal layers
remains roughly constant. In contrast the in-plane $J$s change and 
eventually become antiferromagnetic. In particular the coupling becomes strongly 
antiferromagnetic for $J_{2,1}$, i.e. for those Mn atoms that
get closer in the hexagonal plane under distortion. Also $J_{2,3}$, which couples 
the Mn atoms increasing their separation, is reduced and becomes antiferromagnetic for
large distortions. Finally the coupling parameters $J_{3,1}$ and $J_{3,2}$
have only minor changes, with $J_{3,1}$ becoming weakly
antiferromagnetic for large distortions. 

The evolution of the coupling constants with the distortion
explains why for the $+-+-$ and $+--+$ spin configurations (table
\ref{tab:relaxcell}), where the spins are antiferromagnetically
aligned in the hexagonal plane, the lowest energy is found for the
B31 structure.  In fact the relaxed structure for both spin configurations
is similar to a distortion of about 200 \%. At this distortion the
in-plane coupling constants $J_{2,1}$ and $J_{2,3}$ become
antiferromagnetic, resulting in a reduction of the total energy as
compared to the B$8_1$ structure. 

\begin{figure}
\centering
\subfigure[]{
	\label{fig:tcrel_od}
	\includegraphics[width=7.5cm]{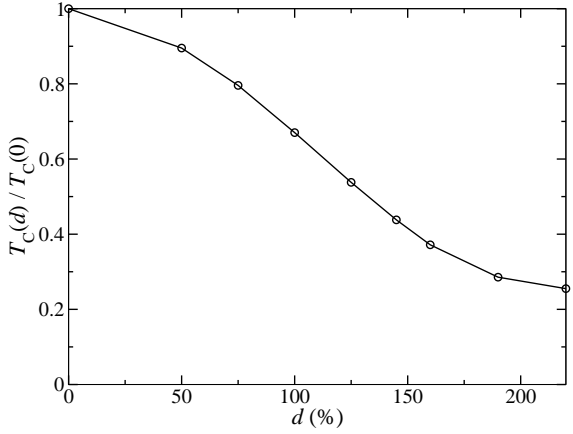}}
\subfigure[]{
	\label{fig:ene_od}
	\includegraphics[width=7.5cm]{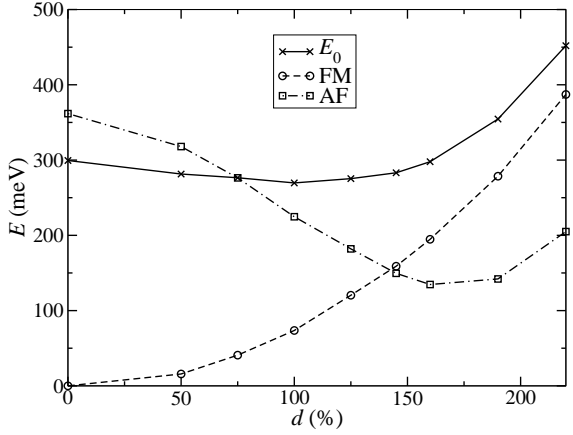}}
\caption{(a) Relative change of the mean field Curie temperature
$(T_\mathrm{C}(d)-T_\mathrm{C}(0))/T_\mathrm{C}(0)$ for the
ferromagnetic state. (b) Total energy for one B31 unit cell for
the ferromagnetic configuration (FM), for the $+--+$
antiferromagnetic configuration (AF), together with $E_0$ (equation
\eqref{eq:heis}), as a function of the distortion $d$. $d=0$ \% and
$d=100$ \% represent respectively the B8$_1$ and the B31 phase at $T_\mathrm{p}\approx$318  K.}
\label{fig:ene_tcrel_od}
\end{figure}

Figure \ref{fig:tcrel_od} shows the relative change of the mean
field Curie temperature $T_\mathrm{C}(d)/T_\mathrm{C}(0)$ for the
ferromagnetic state.  $T_\mathrm{C}$ decreases monotonically with
increasing the distortion. For 100~\% distortion (B31 structure at $T_\mathrm{p}$)
$T_\mathrm{C}(100\%)/T_\mathrm{C}(0)=0.67$, demonstrating that when
the phase transition from the B$8_1$ to the B31 structure occurs,
the system in the B31 cell is already paramagnetic with very
little magnetic order.  The experimental Curie temperature
$T_\mathrm{C}^\mathrm{exp}$ for the hexagonal cell at
$T=T_\mathrm{p}$ is not known, since the structure changes. 

Figure \ref{fig:ene_od} shows the total energy per B31 unit cell as a function 
of the distortion in the ferromagnetic (FM) and in the $+--+$ antiferromagnetic 
configurations (AF). This latter is the antiferromagnetic configuration giving
the lowest total energy at its minimum among all the ones calculated along 
the considered distortion.
The figure also shows the value of $E_0$, the energy of the paramagnetic state 
(see equation (\ref{eq:heis})). The zero in the energy scale is the energy of the 
ferromagnetic state for $d=0$~\%. 

The ferromagnetic state has its energy minimum for $d=0$~\%, and
increases parabolically for increasing distortion. This means that the B8$_1$
structure is the one with lowest energy in the ferromagnetic state. In contrast the
competing antiferromagnetic configuration has a minimum for about 180~\%
distortion, where the energy is lower than the ferromagnetic phase.
The ground state of the system is therefore expected to be ferromagnetic for distortions
up to about 100~\%, and then to become antiferromagnetic for
$d>$180~\%.  For distortions in between the ground state is expected to be a non-collinear, 
canted spin structure. 

$E_0$ has a very flat minimum for distorted cells, reflecting the fact that
the total energy increases for the ferromagnetic state, but
decreases on average for the antiferromagnetic states. The minimum
is found to be at about $d\approx100$~\%, which corresponds indeed
to the lattice parameters of the paramagnetic state above the
phase transition.  This explains the structural change from the
B$8_1$ to the B31 structure when the MnAs magnetic state goes from
ferromagnetic to paramagnetic.  The phase transition at
$T_\mathrm{p}$ can be seen as a jump from the minimum of the FM
curve to the minimum of the $E_0$ curve.

\subsection{B31 to B8$_1$ distortion at $T_\mathrm{t}$}
\label{sec:disttt}

For temperature in between $T_\mathrm{p}$ and $T_\mathrm{t}$ the MnAs crystal structure
continuously changes from B31 to B$8_1$. As mentioned in section \ref{sec:expprop} the phase
transition temperature $T_\mathrm{t}$ is usually identified as the temperature where the 
susceptibility and the specific heat have a maximum. This is at about 398~K.
However, the distortion disappears at slightly higher temperatures as pointed out in 
reference [\onlinecite{zieba1}]. Therefore, since the exact temperature for this second order
structural phase transition is not known exactly, we introduce an operative definition and 
assume that the distortion disappears at a temperature $T_\mathrm{s}$, at which the 
slope of the in-plane lattice constant as function of temperature $a(T)$ changes discontinuously 
(see figure \ref{fig:acofT}). According to figure \ref{fig:acofT} the lattice constant at 
$T_\mathrm{s}$ is $a(T_\mathrm{s})=\tilde{a}\approx3.699$ \AA, and $(\partial
a/\partial T)_{T_\mathrm{s}}\approx0$. At $T_\mathrm{t}$ the same lattice parameter
is $a(T_\mathrm{t})\approx$3.697 \AA, so that the difference in $\tilde{a}$ is very small.

The main reason for the second order phase transition at high temperatures is related 
to the lattice thermal expansion. The idea is that upon volume expansion, the ground state of the
paramagnetic phase moves towards the hexagonal structure. We verify this hypothesis by calculating 
the minimum of $E_0$ ($E_0^\mathrm{min}$) along a distortion of the cell transforming B$8_1$ 
to B31. In the calculation the volume of the cell is kept constant and we repeat the calculation
for different volumes. This allows us to evaluate
both $E_0^\mathrm{min}$ and the corresponding distortion as a function of the volume.
Since for $T >T_\mathrm{p}$ MnAs is always paramagnetic, then the minimum of $E_0$
corresponds to the stable distortion $d_0$ at a given volume. In practice the change in volume
can be described simply by the change in the planar lattice constant $a$, since both 
$b/a$ and $c/a$ do not deviate much from their value at $T_\mathrm{p}$. Thus we
always consider $b=\sqrt{3}~a$ and $c=1.556~a$ and the phase transition is investigated
as a function of $a$ only.

\begin{figure}
\center
\includegraphics[width=7cm,clip=true]{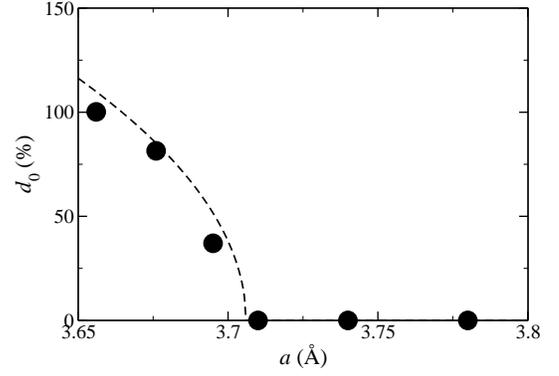}
\caption{Distortion $d_0$ for the minimum of the paramagnetic ground
state as a function of the lattice constant $a$. The dots are
calculated values, the dashed line is a fit with equation
\eqref{eq:dofa}.}
\label{fig:d_of_V}
\end{figure}
The equilibrium distortion $d_0$ as a function of $a$ is presented in figure \ref{fig:d_of_V}.
Indeed the distortion decreases with volume and it disappears for $a$ between $a=3.695$
\AA~and $a=3.71$ \AA. Moreover we find 100\% distortion for $a\sim3.66$ \AA. These values 
agree rather well with the experimental ones, where the distortion disappears at
about $a(T_\mathrm{s})=3.699$ \AA, and 100\% distortion is found
at $a(T_\mathrm{p})=3.673$ \AA. 

In order to interpret these results consider that the distortion is symmetric for $\pm d$ 
($E_0(d)=E_0(-d)$), and therefore $E_0$ can be expanded in even powers of the distortion
$E_0(d)=r_0 + r_1~d^2 + r_2~d^4$. Here $r_i$ are parameters to fit to the
DFT calculations. In particular note that $r_0$ corresponds to the energy of the
paramagnetic phase when the crystal is undistorted, i.e. it has the hexagonal structure. 
In this way the minimum of the $E_0(d)$ curve is obtained for 
$d_0=\sqrt{-\frac{r_1}{2 r_2}}$. For small distortions the parameters $r_1$
and $r_2$ can be further expanded around $\tilde{a}$ as $r_1= r_{1,1} (a-\tilde{a})$ and
$r_2=r_{2,0}+r_{2,1}~(a-\tilde{a})+ r_{2,2}~(a-\tilde{a})^2$. The
calculated values for the leading terms are $\tilde{a}=3.706$ \AA,
$r_{1,1}=62.4~ 10^{-3}$ meV/\AA~and $r_{2,0}=1.29~10^{-7}$ meV,
($d$ is given in percent). The equilibrium distortion $d_0$ up to first
order in $a$ is then
\begin{equation}
d_0(a)=\gamma
\sqrt{\frac{\tilde{a}-a}{\tilde{a}}}~\Theta(\tilde{a}-a),~~~~~~\gamma=\sqrt{\tilde{a}
\frac{r_{1,1}}{2~r_{2,0}}},
\label{eq:dofa}
\end{equation}
where $\Theta(x)$ is the Heaviside function. With the values of $\tilde{a}$,
$r_{1,1}$ and $r_{2,0}$ given above $\gamma=947$ is obtained. The 
resulting distortion is presented in figure \ref{fig:d_of_V}. 

\begin{figure}
\center
\includegraphics[width=7cm,clip=true]{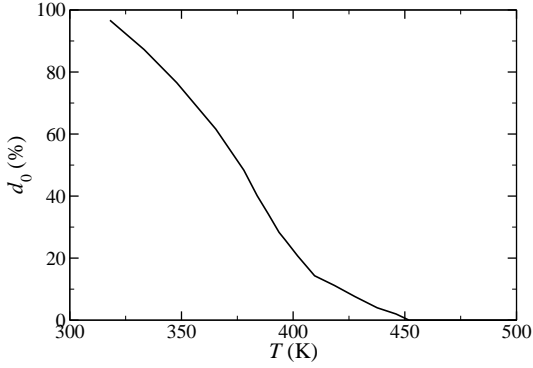}
\caption{Distortion $d_0$ for the minimum of the paramagnetic
ground state as a function of temperature calculated with equation
\eqref{eq:dofa} (with $\gamma=1184$ and $\tilde{a}=3.699$) using
$a(T)$ taken from figure \ref{fig:acofT}.}
\label{fig:d_of_T}
\end{figure}

Interestingly if we use the equation (\ref{eq:dofa}) to fit the experimentally
determined distortions at $a(T_\mathrm{s})=3.699$\AA\ ($d_0=100$\%) and
$a(T_\mathrm{p})=3.673$\AA\ ($d_0=0$), we obtain $\gamma=1184$ and 
$\tilde{a}=3.699$, both in good agreement with our calculated values.  
This suggests that the main effects of the distortion to the B31 structure arise from
the atomic displacement from the symmetry positions, and that small changes of the 
ratio of the lattice vectors, neglected in our calculations, play only a secondary role.  
Using the values $\gamma=1184$ and $\tilde{a}=3.699$ of the two parameters the
evolution of the distortion as a function of temperature $T$ can
be obtained by inserting the data for $a(T)$ from figure
\ref{fig:acofT} in equation \eqref{eq:dofa}.  The result is shown
in figure \ref{fig:d_of_T}. It agrees well with the experimental
result (see figure 3 in reference \onlinecite{suzuki1}).  The main
difference however is that in reference \onlinecite{suzuki1} the distortion 
$d_0$ becomes zero at 398 K, whereas in our results this happens only
at 450 K. This is due to our choice of
$T_\mathrm{s}=450$ K, which by definition sets the temperature
where the distortion disappears.  Close to the phase transition
temperature fluctuations play an important role, so that for very
small distortions close to the phase transition the description
may not be valid.

\begin{figure}
\center
\includegraphics[width=7cm,clip=true]{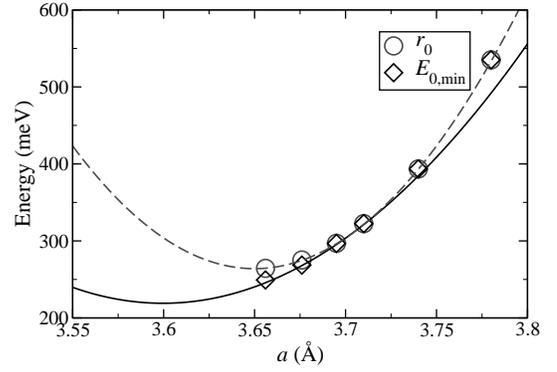}
\caption{Energy expansion coefficient $r_0$ and $E_\textrm{0,min}$
as function of the lattice constant. The
dashed line shows $\epsilon_0+ \epsilon_1~(a-a_0)^2$, the
solid line shows $\epsilon_0+ \epsilon_1~(a-a_0)^2  -
\epsilon_2~ (a-\tilde{a})^2$ (see equation \eqref{eq:eofa}).}
\label{fig:E0_of_V}
\end{figure}

By using the computed values of $r_0$, $r_1$ and $r_2$ the minimum of
$E_0$ ($E^\textrm{min}_0$=$E_0(d=d_0)$) is calculated as a function of the volume of the
unit cell, and it is shown in figure \ref{fig:E0_of_V} together with $r_0$. Recalling that $r_0$
is the energy of the paramagnetic hexagonal phase, it also can be expanded as
function of the lattice constant $r_0=\epsilon_0+
\epsilon_1~(a-a_0)^2$, where $a_0$ is the equilibrium lattice constant of the hexagonal phase. 
This, combined with equation \eqref{eq:dofa},
gives an expression for the energy minimum as a function of the lattice
constant $a$
\begin{equation}
\begin{split}
E^\textrm{min}_0(a)&=\epsilon_0+ \epsilon_1~(a-a_0)^2  - \epsilon_2~
(a-\tilde{a})^2~\Theta(\tilde{a}-a),\\
\epsilon_2&=\frac{r_{1,1}^2}{4 r_{2,0}}=7546~
\frac{\textrm{meV}}{\textrm{\AA}^2},
\end{split}
\label{eq:eofa}
\end{equation}
where $\epsilon_0, \epsilon_1$ and $a_0$ are to be fitted from the
calculations of $r_0$ (figure \ref{fig:E0_of_V}). The fitted
values are $\epsilon_0=264$ meV, $\epsilon_1=15935$ meV/\AA$^2$
and $a_0=3.65$ \AA. From the equation (\ref{eq:eofa}) the energy minimum is easily
found
\begin{equation}
a_\mathrm{min}=a_0~\left( 1 -
\frac{\epsilon_2}{\epsilon_1-\epsilon_2}\frac{\tilde{a}-a_0}{a_0}\right)\:,
\end{equation}
and by using the calculated parameters we estimate $a_\mathrm{min}=3.60$
\AA.  Since $a_\mathrm{min}<a_0$, we derive the important result that the distortion 
allows the volume to be further reduced as compared to the hexagonal phase. 
Furthermore the curvature of the energy as a function of $a$ is
\begin{equation}
\frac{\partial^2 E^\textrm{min}_0(a)}{\partial
a^2}=2\epsilon_1-2\epsilon_2~\Theta\left(\tilde{a}-a\right),
\end{equation}
which is also reduced by a factor $2\epsilon_2$ when the structure is
distorted.  

The effect of the thermal expansion on the lattice parameter can now be
modeled as a temperature dependent position of $a_0$, $a_0=a_0(T)$. The
change of the lattice constant with temperature for the
distorted phase can therefore be written as
\begin{equation}
\frac{\partial a_\mathrm{min}}{\partial
T}=\frac{\epsilon_1}{\epsilon_1-\epsilon_2}\frac{\partial
a_0}{\partial T}
\end{equation}
Since $\epsilon_2$ is smaller than $\epsilon_1$ the lattice
expands with temperature faster for the distorted phase than for
the undistorted phase. If the calculated values for $\epsilon_1$
and $\epsilon_2$ are used the ratio
$\epsilon_1/(\epsilon_1-\epsilon_2)$ is found to be 1.90, which agrees well with
the value of 2.25 extracted from figure \ref{fig:acofT}. Near the
phase transition, i.e. where $d\approx0$, phononic effects due to
the different curvatures of the energy and fluctuations should be
considered. It is especially interesting that the change of the
lattice constant with the temperature goes to zero near the phase
transition temperature.

In the same way as $E_0$ also $T_\mathrm{C}$ can be expanded as a
function of the lattice constant and of the distortion
\begin{equation} 
T_\mathrm{C}(a,d)=T_\mathrm{C}(\tilde{a})~\left[1+K_{v}~
\frac{a-\tilde{a}}{\tilde{a}} -K_{d}~d^2\right], 
\label{eq:tcofa_m1}
\end{equation} 
where  $K_{v}$ and $K_{d}$ are parameters, and
$T_\mathrm{C}(\tilde{a})$ is the Curie temperature for the cell
with lattice parameter $\tilde{a}$ and where the atoms are in the
hexagonal positions.  The parameters are fitted by least mean
squares to the calculated values of $T_\mathrm{C}$ obtained for
six lattice constants ranging between 3.656 \AA~and 3.78~\AA\ and
for different distortions.
We obtain $T_\mathrm{C}(\tilde{a})$=573 K, $K_{v}=6.80$ and
$K_{d}=2.62~10^{-5}$.  At $d=100~\%$ the relative change of the
Curie temperature is $T_\mathrm{C}(\tilde{a},d=100\%)/T_\mathrm{C}(\tilde{a})=0.74$ and
corresponds roughly to the value of figure \ref{fig:tcrel_od}. In this case
it is slightly larger due to the fact that the
volume is kept constant, whereas for the calculations of figure \ref{fig:tcrel_od}
it shrinks with increasing the distortion.

Next we calculate the dependence of the magnetic moment on the
distortion and on the unit cell volume. The dependence is
again expanded to lowest order in $a$ and $d$
\begin{equation}
\begin{split}
\mu(a)=\mu(\tilde{a})\left[1+\alpha_{\mu}~\frac{a-\tilde{a}}{\tilde{a}}
+ \alpha_{\mu,d}~d^2\right]\;.
\label{eq:muofa}
\end{split}
\end{equation}
We now have different ways of extracting the magnetic moment of the Mn atoms 
from our DFT calculations. One possibility is to take the total
moment of the cell for the ferromagnetic spin configuration and
divide it by the number of Mn atoms. In this way however the small induced
moments of the As atoms are subtracted from the moment on the Mn. 
A second possibility is to take the average Mulliken spin
population for the Mn atoms. The advantage of this method is that
also antiferromagnetic configurations can be used to determine the
average moment, and the induced moments of the As atoms are
accounted for. The drawback however is that Mulliken populations are
somewhat arbitrary as they depend on the basis set. 

By setting $\tilde{a}$ to 3.699 \AA, the values obtained using the
cell moment are $\mu(\tilde{a})=3.28~\mu_\mathrm{B}$,
$\alpha_{\mu}$=3.28 and $\alpha_{\mu,d}$=$1.15~10^{-6}$. Similarly from 
the average Mulliken population over all the magnetic configurations
we obtain $\mu(\tilde{a})$=3.42 $\mu_\mathrm{B}$,
$\alpha_{\mu}$=3.48, and $\alpha_{\mu,d}\approx 0$. These
results are rather similar to each other. Our analysis
shows that the change of the magnetic moment is mainly due to
the change in volume, whereas the distortion has basically no
effect. 

\subsection{Small distortions of the B8$_1$ structure}
\label{smalldist}

In this section the dependence of $T_\mathrm{C}$ on the individual
lattice parameters and on the distances between the atoms is
investigated for the B8$_1$ structure.  Our approach is to distort
the cell orthorhombically but to leave the
atoms in their high symmetry positions.  Apart from a
general understanding of the phase diagram of MnAs this analysis
is useful for predicting the behavior of MnAs when
grown on a substrate.  For instance when grown on GaAs(001) the substrate 
induces strain in MnAs, and the unit cell is slightly orthorhombically distorted.
\cite{ploog1} This distortion does not correspond to the
orthorhombic B31 structure, since the atoms do not move out of the
high symmetry positions.  Moreover different growth orientations are
possible, and the Curie temperature varies accordingly.
\cite{iikawa1} In addition reference [\onlinecite{mnas1}] presents 
experimental results showing that the phase transition
temperature $T_\mathrm{p}$ changes when strain is applied to the
MnAs film.  In that article we have compared our theoretical
predictions for the dependence of $T_\mathrm{p}$ on the lattice
distortion with the experimental findings. In this section we refine and
expand our previous analysis.

When the cell is orthorhombically distorted the first three
nearest neighbor coupling constants split into five different
constants, corresponding to the ones of the B31 cell
(figure \ref{fig:js}), and with the only exception that now
$J_{2,1}=J_{2,3}$. The number of total energy calculations for the
fit of the coupling parameters is 16.
\begin{figure}
\includegraphics[width=7.5cm,clip=true]{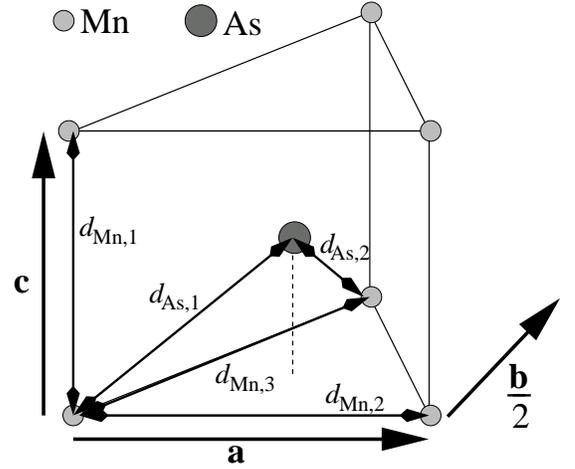}
\caption{Diagram of the positions of Mn and As atoms in
orthorhombic MnAs.}
\label{figure3}
\end{figure}
The change of $T_\mathrm{C}$ for each different distortion can be
expressed as a function of the change of each single Mn-Mn and
Mn-As distance in the unit cell. This gives
\begin{equation}
\frac{\delta T_\mathrm{C}}{T_\mathrm{C}}=\frac{\delta J_0}{J_0} = 
\sum_{\mu=1}^5 K_{\mu}N_{\mu}~ \frac{\delta d_{\mu}}{d_{\mu}},
\label{eq:Kvald}
\end{equation}
where the sum goes over all five independent distances in the
orthorhombic unit cell as defined in figure~\ref{figure3}. 
The dependence on the angles between the atoms
is neglected. $N_\mu$ are the multiplicities of each distance
$d_\mu$ within one unit cell, and have the values
$N_\mathrm{Mn,1}=4,~N_\mathrm{Mn,2}=4,
N_\mathrm{Mn,3}=8,~N_\mathrm{As,1}=4,~N_\mathrm{As,2}=8$.  
For the evaluation of the coefficients $K_\mu$, 21 different distortions are considered, 
including changes of volume, changes of the ratio of the different axes and different
displacements of the As atoms. For all the distortions the
orthorhombic symmetry however was preserved.  The best fit for $K_\mu$
gives
\begin{equation}
\begin{split}
K_\mathrm{Mn,1}&= 6.6,~ K_\mathrm{Mn,2}=6.2,~ K_\mathrm{Mn,3}=5.9,\\
K_\mathrm{As,1}&= -7.5,~ K_\mathrm{As,2}=-7.5.
\end{split}
\end{equation}
The values of $K_\mathrm{Mn,1}$ and $K_\mathrm{Mn,2}$ are almost
identical, as one should expect from the symmetry and similarly for
$K_\mathrm{As,1}$ and $K_\mathrm{As,2}$. 

Equation \eqref{eq:Kvald} describes the fact that the change of
$T_\mathrm{C}$ is the result of an interplay between the change of
the Mn-Mn and Mn-As distances. The calculated $K_\mu$ show that, 
while an increase in the distance between Mn atoms increases
$T_\mathrm{C}$, an increase of the Mn-As distance
decreases it. However note that the two distances can not be
changed independently, thus the net change in $T_\mathrm{C}$
depends on the details of the distortion.

$J_0$ can also be expanded over the orthorhombic lattice
parameters
\begin{equation}
\frac{\delta J_0}{J_0} = \sum_{i=1}^3~K_i\frac{\delta
a_{i}}{a_{i}},~\textrm{with}~K_i=\sum_{\mu=1}^5 K_{\mu}N_{\mu}~
\frac{a_i}{d_\mu}\frac{\partial d_\mu}{\partial a_i},
\label{eq:j0lin}
\end{equation}
where $a_1=a$, $a_2=b$ and $a_3=c$.  The change of the distances
between Mn and As atoms is not exactly known for the orthorhombic
cell. However it is easy to show that to first order the position
of the As atom in the cell does not influence $J_0$, since up to
first order $\delta d_\mathrm{As,2}=2 \delta d_\mathrm{As,1}$ when
moving the As atom inside the cell.  Therefore the As atoms can
be assumed to remain in the high symmetry position.  Assuming now
$K_2=K_3$ and $K_4=K_5$ (as imposed by symmetry), the general
form of the $K_i$ is
\begin{equation}
\begin{split}
K_a=K_b&=6K_2+\frac{96K_4}{16+3~\tilde{c}^2},\\
K_c&=4 K_1+\frac{36 K_4~\tilde{c}^2}{16+3~\tilde{c}^2},
\label{eq:j0ac}
\end{split}
\end{equation}
where $\tilde{c}=c/a$. Using the average between $K_2$ and $K_3$ and
$\tilde{c}=1.533$ this gives
\begin{equation}
K_a=K_b=4.9,~~ K_c=-1.0 ~.
\label{eq:Ks}
\end{equation}

Our results clearly show that stretching the unit
cell along the basal plane raises $T_\mathrm{p}$ (since $K_a~>~
0$), while stretching along the $c$-axis lowers $T_\mathrm{p}$
($K_c~<$~0).  An increase of the volume without distorting the
cell results in an increase of the ferromagnetic exchange
interactions and therefore of $T_\mathrm{C}$, since $K_a$ is
positive and larger in magnitude than $K_c$. If the cell changes
only its volume the expansion corresponds to the one of equation
\eqref{eq:tcofa_m1} with a factor $K_v=2 K_a+K_c=8.8$.
These results differ slightly from the ones given in reference
[\onlinecite{mnas1}], since in that case we did not constrain $K_a$ to be
equal to $K_b$. 

\section{Discussion}

\subsection{Curie temperature and susceptibility}
\label{susceptibility}

It is now possible to analyze two peculiar properties of MnAs.
The first is the anomalous behavior of the susceptibility $\chi$
as a function of temperature between $T_\mathrm{p}$ and
$T_\mathrm{t}$. The second is the fact that although the Curie temperature
for ferromagnetic MnAs has to be larger than $T_\mathrm{p}$,
the $T_\mathrm{C}$ extrapolated from the susceptibility
above $T_\mathrm{t}$ is only 285 K. \cite{goodek1} In this section
both these features are explained using the dependence of the
susceptibility on the Curie temperature ($\chi=\chi(T_\mathrm{C})$), and 
the strong dependence of the Curie temperature on the lattice parameters 
$T_\mathrm{C}=T_\mathrm{C}(a,b,c,d)$.
$T_\mathrm{C}(T)$ and $\chi(T)$ are therefore determined using the
experimentally measured temperature dependence of the lattice
vectors $a(T), b(T)$ and $c(T)$ and of the
distortion $d(T)$. This analysis also provides a tool for extracting
the parameters $K_v$ and $K_d$ from experimental data.

By generalizing the equations \eqref{eq:tcofa_m1} and \eqref{eq:j0lin}
the Curie temperature $T_\mathrm{C}(a,b,c,d)$ can be written as
\begin{equation}
\begin{split}
&T_\mathrm{C}(a,b,c,d)=\\
&~~T_\mathrm{C,0}\left[1+ K_a \left( \frac{a-\tilde{a}}{\tilde{a}}+
\frac{b-\tilde{b}}{\tilde{b}}\right) +  K_c ~ \frac{c-\tilde{c}}{\tilde{c}} - K_{d}~d^2\right],
\end{split}
\label{eq:tcofadist}
\end{equation}
where we use the fact that $K_a=K_b$. We take the values for $K_a$ and $K_c$ 
from equation \eqref{eq:Ks}, $K_{d}=2.62~10^{-5}$ and
$T_\mathrm{C,0}=T_\mathrm{C}(\tilde{a})=573$ K are those calculated in
section \ref{sec:disttt}. The reference lattice parameters are
chosen to be the lattice vectors at $T_\mathrm{s}$
($\tilde{a}=3.699~
$\AA$,\tilde{b}=\sqrt{3}~\tilde{a},\tilde{c}=1.56~\tilde{a})$.

Similarly to the Curie temperature also the susceptibility 
is calculated in the mean field approximation. This is justified for $T \gg T_\mathrm{C}$, 
a condition which is satisfied for paramagnetic MnAs.  
The molar susceptibility $\chi_\mathrm{M}$ is then given by 
\begin{equation}
\begin{split}
\chi_\mathrm{M}^{-1}&=\frac{1}{C_0}\left(T-T_\mathrm{C}\right),\\
C_0&=\frac{N_\mathrm{A} \mu_\mathrm{B}^2 g^2}{3 k_\mathrm{B}}
s \left(s+1\right).
\end{split}
\label{eq:chiinv}
\end{equation}
$N_\mathrm{A}$ is the Avogadro's number, $g\approx2$ is the Lande's
factor for the free electron spin, $k_\mathrm{B}$ is the Boltzmann
constant and $s$ is the atomic total spin.  Note that the
susceptibility has an additional implicit temperature dependence since
$T_\mathrm{C}$ and $s$ depend on the temperature through the
lattice distortion. However in what follows we neglect the
dependence of $s$ on the lattice parameters so that $C_0$ is constant over all temperature. 
An analysis performed by relaxing this approximation
gives similar results.

Similarly to section \ref{sec:disttt} the model is further simplified by 
assuming that $b/a$ and $c/a$ are constant above $T_\mathrm{p}$.  
As indicated in equation \eqref{eq:dofa} $d$ then is a function of the
lattice constant, $d(a)=\gamma\sqrt{(\tilde{a}-a)/\tilde{a}}~\Theta(\tilde{a}-a)$. 
Moreover we have shown that the experimental distortion as function of the lattice 
constant is well reproduced when $\gamma=1184$.
Therefore it is now possible to express $T_\mathrm{C}$ and $\chi_\mathrm{M}^{-1}$ 
as a function of the lattice constant $a$ only
\begin{equation}
\begin{split}
&\chi_\mathrm{M}^{-1}(a)=\\
&~~\frac{1}{C_0}
\left[T-T_\mathrm{C,0}\left(1+ \left( K_{v} + K_{d}~\gamma^2
\Theta\left(\tilde{a}-a\right)\right)\frac{a-\tilde{a}}{\tilde{a}}\right)\right],
\label{eq:susc_of_a}
\end{split}
\end{equation}
where $K_{v}=2 K_a + K_c=8.8$. This equation shows that if the
lattice expands strongly with temperature $\chi_\mathrm{M}^{-1}$ decreases.

For temperatures between $T_\mathrm{p}$ and about 390 K, as well
as above $T_\mathrm{s}$, $a$ increases approximately linearly with temperature 
(see figure \ref{fig:acofT}) and can therefore be written as
\begin{equation}
a(T)=a(T_0)\left[1+\alpha~\frac{T-T_0}{T_0}\right],
\label{eq:aoft_lin}
\end{equation}
where the experimental values for the coefficients are
$T_{0,+}=T_\mathrm{s}=452$ K, $a(T_\mathrm{s})=\tilde{a}=3.699$
\AA ~and $\alpha_{+}=0.0126$ for temperatures above $T_\mathrm{s}$
(the index ``$+$'' denotes the high temperature region above
$T_\mathrm{s}$), and $T_{0,-}=T_\mathrm{p}=318$ K,
$a(T_\mathrm{p})=3.673$ and $\alpha_{-}=0.0284$ for temperatures
between $T_\mathrm{p}$ and about 390 K (the index ``$-$'' denotes the
intermediate temperature region). By inserting equation
\eqref{eq:aoft_lin} into equation \eqref{eq:susc_of_a} we obtain
for the high temperature region above $T_\mathrm{s}$
\begin{equation}
\chi_\mathrm{M}^{-1}(T)=\frac{1}{C_\mathrm{eff}}(T-T_\mathrm{C,eff}),
\end{equation}
with 
\begin{equation}
\begin{split}
C_\mathrm{eff}=&\frac{1}{1-K_{v}~\alpha_{+}
\frac{T_\mathrm{C,0}}{T_\mathrm{s}}}~C_0,\\
T_\mathrm{C,eff}=&\frac{ 1+ K_{v} \alpha_{+}} {1-K_{v}
\alpha_{+}  \frac{T_\mathrm{C,0}}{T_\mathrm{s}}}~T_\mathrm{C,0}.
\end{split}
\end{equation}
$T_\mathrm{C,eff}$ and $C_\mathrm{eff}$ are the experimentally
accessible quantities for the high temperature susceptibility, and
due to the expansion of the lattice they are different from
$T_\mathrm{C,0}$ and $C_0$. The experimentally measured values are
$T_\mathrm{C,eff}$= 285 K and $C_\mathrm{eff}=3.12~10^{-5}$ m$^3$
K, \cite{goodek2} which corresponds to an effective magnetic
moment of 3.57 $\mu_\mathrm{B}$.  From $C_\mathrm{eff}$ and
$T_\mathrm{C,eff}$ the values of
$T_\mathrm{C,0}=T_\mathrm{C}(\tilde{a})$ and $C_0$ can now be
obtained
\begin{equation}
\begin{split}
C_0=&\frac{1-K_{v} \alpha_{+}}{1-K_{v}
\alpha_{+}\left(1-\frac{T_\mathrm{C,eff}}{T_\mathrm{s}}\right)}~C_\mathrm{eff},\\
T_\mathrm{C,0}=&\frac{1}{1-K_{v}
\alpha_{+}\left(1-\frac{T_\mathrm{C,eff}}{T_\mathrm{s}}\right)}~T_\mathrm{C,eff}.
\label{eq:invct}
\end{split}
\end{equation}
All the variables on the right hand side of equation
\eqref{eq:invct} can be obtained from experiment except $K_v$. For
small $K_v$ the difference between $T_\mathrm{C,0}$ and
$T_\mathrm{C,eff}$ is proportional to $K_v$. Since $T_s$ is larger
than the experimental value of $T_\mathrm{C,eff}$ the effect of
the thermal expansion of the hexagonal structure is a reduction of
the slope of the inverse susceptibility as a function of
temperature, as well as a reduction of the extrapolated Curie
temperature as compared to the real Curie temperature.

In the region where  linear expansion holds the slope of the inverse
susceptibility above $T_\mathrm{t}$ is
\begin{equation}
\frac{\partial{\chi_\mathrm{M,+}^{-1}}}{\partial T}=\frac{1}{C_0}\left(1- K_{v}~
\alpha_{+} \frac{T_\mathrm{C,0}}{T_\mathrm{t}}\right),
\label{eq:chip}
\end{equation}
whereas for in the intermediate temperature region above
$T_\mathrm{p}$ it is
\begin{equation}
\frac{\partial{\chi_\mathrm{M,-}^{-1}}}{\partial T}=\frac{1}{C_0}~\left(1- \left(K_{v}
+ K_{d}~\gamma^2 \right)~
\alpha_{-}\frac{a_0}{\tilde{a}} \frac{T_\mathrm{C,0}}{T_{0,-}}\right).
\label{eq:chim}
\end{equation}
In both regions there is a reduction of the slope due to the
expansion of the lattice. However the reduction is much larger 
for $\chi_\mathrm{M,-}$ than for $\chi_\mathrm{M,+}$, since there
is the additional term proportional to $K_d$ due to the distortion, 
and also $\alpha_->\alpha_+$. As a rough approximation it can be assumed that
$(a_0~T_\mathrm{C,0})/(\tilde{a}~ T_\mathrm{0,-})\approx1$, so
that $\frac{\partial{\chi_\mathrm{M,-}^{-1}}}{\partial T}$ becomes
negative for
\begin{equation}
\left(K_{v}+ K_{d}~\gamma^2\right)\alpha_{-}>1~.
\end{equation}
The values of $\alpha$ and $\gamma$ are determined experimentally
and describe how the structure changes with temperature, whereas
$K_v$ and $K_d$ describe how $T_\mathrm{C}$ varies for
distorted cells. By using our calculated values for $K_v$ and $K_d$ we
obtain $\left(K_{v}+ K_{d}~\gamma^2\right)\alpha_{-}=1.29$. This
is indeed larger than one. Therefore we do predict a negative slope for the 
inverse susceptibility in the intermediate temperature region.  

Finally we extract the values for $K_{v}$ and $K_{d}$ from the experimental 
behavior of the Curie temperature. Since the ratio between $K_a$ and $K_c$
can not be obtained from the thermal properties of MnAs, it is
therefore assumed that $K_c/K_a=-1/4.9=-0.2$ is fixed and corresponds to our
calculated value. For the hexagonal cell ($d=0$) the equation \eqref{eq:tcofadist} reads
\begin{equation}
T_\mathrm{C}(a,c)=T_\mathrm{C,0}\left[1+ K_v\left(k_a \frac{a-\tilde{a}}{\tilde{a}}
+ k_c ~ \frac{c-\tilde{c}}{\tilde{c}} \right)
\right],
\end{equation}
where $k_a=2/(2+K_c/K_a)=1.11$ and $k_c=(K_c/K_a)/(2+K_c/K_a)=-0.11$.  
This has to be valid for all temperatures where the cell is hexagonal, therefore it can
not be assumed that the ratio between $c$ and $a$ is constant
since it changes abruptly from 1.533 to 1.556 at $T_\mathrm{p}$.
By using the expression for $T_\mathrm{C,0}$ from equation \eqref{eq:invct} 
we obtain $K_v$
\begin{equation}
K_v=\frac{\frac{T_\mathrm{C}(a,c)}{T_\mathrm{C,eff}}-1}{\alpha_+ 
\left(\frac{T_\mathrm{C}(a,c)}{T_\mathrm{C,eff}} - \frac{T_\mathrm{C}(a,c)}{T_\mathrm{s}}\right)
+ k_a ~ \frac{a-\tilde{a}}{\tilde{a}}
+ k_c ~ \frac{c-\tilde{c}}{\tilde{c}} }.
\label{eq:kv}
\end{equation}
As reference Curie temperature the extrapolated value to room
temperature is used, which can be estimated to be about
$T_\mathrm{C}(a=3.724$ \AA, $c=1.533 a)$=360 K. By inserting the
experimental values for the parameters on the right hand side of
the equation (\ref{eq:kv}) we obtain $K_v=18.1$. This is
about twice as big as our predicted value. The disagreement may be
due to the fact that the ratio between $K_c$ and $K_a$ has been
fixed for our calculated value.  By inserting this value for $K_v$ in equation
\eqref{eq:invct} we obtain $T_\mathrm{C,0}=311$ K, the Curie
temperature for the lattice parameters at $T=T_\mathrm{s}$.

In order to extract $K_d$ from experiments we use the relative change in
the slope of the inverse susceptibility around $T_t$.  This is, according to the
equations \eqref{eq:chip} and \eqref{eq:chim},
\begin{equation}
\frac{\frac{\partial{\chi_\mathrm{M,+}^{-1}}}{\partial
T}-\frac{\partial{\chi_\mathrm{M,-}^{-1}}}{\partial
T}}{\frac{\partial{\chi_\mathrm{M,+}^{-1}}}{\partial
T}}=1 - \frac{1-\left(K_{v} + K_{d}~\gamma^2 \right)~
\alpha_{-}\frac{a_0}{\tilde{a}} \frac{T_\mathrm{C,0}}{T_0}}{1- K_{v}~
\alpha_{+} \frac{T_\mathrm{C,0}}{T_\mathrm{t}}},
\label{eq:slope}
\end{equation}
and increases with increasing $K_v$ and $K_d$. All the variables
in this equation can be derived from the experimental measurements, except
$K_v$ and $K_d$. Experimentally different values are found for the
relative change of the slope (left hand side of equation (\ref{eq:slope})).
\cite{zieba1,guillaud1,basinski1,selte2} These are all of the order of
$1.44$. By using this value for the relative change of the slope and
the previously calculated value $K_v=18.1$, $K_d$ is found to be
$1.78~10^{-5}$. This value agrees approximately with our
predicted value of $2.62~10^{-5}$.  

In conclusion the table \ref{tab:paramdftexp} summarizes the parameters calculated
in this work, by comparing our \textit{ab initio} results obtained from the DFT calculations 
and the Heisenberg model, with the results obtained by fitting to the experimental data.
In general there is a fair agreement, although the the DFT results underestimate $K_v$ and overestimate
$K_d$.

%
\begin{table}[thb] 
\center
\begin{tabular}{p{1.6cm}p{0.3cm}*{8}{p{0.8cm}}}
\hline\hline\\
                    & & $T_{C,0}$ (K)& $K_d$  (~10$^{-5}$) & $K_v$              & $K_a$ & $K_c$  & $\gamma$           & $\tilde{a}$ (\AA)\\
\hline\\                                                                              
  DFT               & &  579         &  2.62               & \hspace{0.15cm}8.8 & 4.9   & -1.0   & \hspace{0.15cm}947 & 3.706            \\
  FIT & &  311         &  1.78               & 18.1               & 10.1  & -2.1   & 1184               & 3.699            \\
\hline\hline
\end{tabular}
\caption{\label{tab:paramdftexp}Main parameters used in the description of the phase diagram of
MnAs. We compared results obtained from \textit{ab initio} calculations and Heisenberg model (DFT),
with those of the best fit to the experimental properties (FIT).}
\end{table}

\subsection{Phase diagram} \label{phase diagram}

It is now possible to draw a qualitative description of the  phase
diagram of MnAs (see figure \ref{fig:phased}). Since the magnetic
interactions are strongly dependent on the volume, the effects of
the thermal expansion of the lattice are important.  First the
state of MnAs at zero pressure is considered. For $P=0$
and temperatures below 155~K both ferromagnetic and canted spin
structures can be stabilized. The ground state is the ferromagnetic B8$_1$ 
structure, however also the B31 structure is stable, because at these low 
temperatures the thermal energy is not large enough to induce 
a phase transition from a canted spin structure to a ferromagnetic one. 
As the temperature increases above 155~K the thermal fluctuation may 
induce the phase transition, and therefore only the ferromagnetic state survives.

As the temperature is further increased and the system becomes
paramagnetic, the B31 structure becomes the stable state, as described
in section \ref{disttp} (figure \ref{fig:ene_od}). MnAs then remains
paramagnetic for all temperatures, but because of the
thermal expansion the orthogonal distortion
decreases, until it vanishes when a critical volume is
reached (see figure \ref{fig:d_of_V}).  Then the volume becomes large
enough to stabilize the B$8_1$ structure even for a paramagnetic
state, leading to the second structural phase transition at
$T_\mathrm{t}$.

As pressure is applied to the system the volume is reduced and
thus the ferromagnetic exchange interaction decreases. Therefore
$T_\mathrm{p}$ decreases with pressure, while $T_\mathrm{t}$
increases. At a temperature of approximately 230~K
the ferromagnetic state is stable up to a pressure of 2kbar.  For
high pressures above 2.5-4~kbar the volume is small enough that
canted spin structures become the ground state. The temperature at
which these states become paramagnetic is approximately 230 K. It
increases slightly with pressure, reflecting the fact that the
antiferromagnetic interaction increases as the volume is reduced.

\section{Conclusions} \label{conclusions}

We have investigated by means of \textit{ab initio} electronic structure
calculations the magneto-structural properties of MnAs. The stable structure 
for the ferromagnetic state is found to be the B8$_1$ structure. However
if antiferromagnetic alignment in the hexagonal plane is imposed
the B31 structure becomes more stable. By fitting
the DFT total energies of different magnetic configurations to a
Heisenberg type energy it is shown that the main contributions to
the physical properties originate from the exchange coupling
parameters up to 3rd nearest neighbor. The Curie temperature was
calculated in the mean field approximation, with values approximately twice 
as large as the experimental ones.

The main assumption of the phenomenological model of Bean and
Rodbell \cite{beanrodb2} that the ferromagnetic exchange coupling
parameters increase when the volume is increased has been
confirmed (equation \eqref{eq:Ks}) using this analysis. However it
has been shown that the exchange interactions depend not only on
the volume, but that the orthogonal distortion to the B31
structure plays an important role. For the experimentally observed
distortions some of the in-plane exchange coupling coefficients
become antiferromagnetic.  This is the reason for the stability of
the B31 structure for those configurations of the magnetic moments
that have an antiferromagnetic component in the hexagonal plane.
Different canted spin structures are expected to minimize the
energy for different distortions, since there are both positive and
negative exchange coefficients depending on the amount of
distortion of the B31 structure. 

Furthermore it has been shown that for paramagnetic states the B31 structure
is stable at small volumes, while the B$8_1$ structure is stable
above a critical lattice constant of about 3.7 \AA. This explains
the second order phase transition at $T_\mathrm{t}$, since at that
temperature the lattice constant crosses this critical value.

The Curie temperature has been expanded as a function of the
lattice vectors and of the amount of distortion. An increase in the
volume leads to an enhancement of the Curie temperature, while an
increase of the distortion leads to a reduction.  With these results 
in hand the increase of the susceptibility between $T_\mathrm{p}$ and $T_\mathrm{t}$ has been
explained as the result of the increase of the Curie temperature due to the
change of the structure from the B31 to the B$8_1$ and to the
increase of the volume.  By using the experimental variation of the
lattice parameters with increasing temperature the susceptibility
is indeed found to increase between  $T_\mathrm{p}$ and
$T_\mathrm{t}$. 

A fit of the dependence of the Curie temperature on the lattice
parameters to best reproduce the experimental behavior is also
given.  The calculated values agree within a factor of two with
the values obtained by the \textit{ab initio} calculations. Our
results are in agreement with the phenomenological models
based on the Bean Rodbell idea. But now the various parameters used are
derived from first principles and therefore
validated.


\acknowledgments{We acknowledge the financial support from SFI
(grant number SFI02/IN1/I175).}

\end{document}